\documentclass[aps,pra,superscriptaddress,twocolumn
]{revtex4-2}

\usepackage{graphicx}
\usepackage{amsmath}
\usepackage{amssymb}
\usepackage{amsfonts}
\usepackage{amsthm}
\usepackage{hyperref}
\usepackage{color}
\usepackage{xcolor}
\usepackage{soul}
\usepackage{bm}
\usepackage{dsfont}
\usepackage{bbm}
\usepackage{braket}
\usepackage{enumitem}   
\usepackage{wrapfig}
\usepackage{cleveref}

\usepackage{physics}

\usepackage[draft]{changes}
\usepackage{comment}

\renewcommand{\Tr}[1]{\text{Tr}\left[#1\right]} 

\newcommand{\J}{\mathbb{J}}
\newcommand{\LL}{\mathbb{L}}
\newcommand{\RR}{\mathbb{R}}
\newcommand{\F}{\mathcal{F}}
\newcommand{\D}{{\rm d}}
\newcommand{\hc}{H^{\rm C}}
\newcommand{\hp}{H^{\rm P}}

\newcommand{\eye}{\mathbbm{1}}

\begin{document}

\author{Paolo Abiuso}
\email{paolo.abiuso@oeaw.ac.at}
\affiliation{Institute for Quantum Optics and Quantum Information - IQOQI Vienna,
Austrian Academy of Sciences, Boltzmanngasse 3, A-1090 Vienna, Austria}

\author{Pavel Sekatski}
\email{pavel.sekatski@gmail.com}
\affiliation{D\'{e}partement de Physique Appliqu\'{e}e,  Universit\'{e} de Gen\`{e}ve,  1211 Gen\`{e}ve,  Switzerland}

\author{John Calsamiglia}
\email{john.calsamiglia@uab.cat}
\affiliation{F\'isica Te\`orica: Informaci\'o i Fen\`omens Qu\`antics, Department de F\'isica, Universitat Aut\`onoma de Barcelona, 08193 Bellaterra (Barcelona), Spain}

\author{Martí Perarnau-Llobet}
\email{marti.perarnau@uab.cat}
\affiliation{F\'isica Te\`orica: Informaci\'o i Fen\`omens Qu\`antics, Department de F\'isica, Universitat Aut\`onoma de Barcelona, 08193 Bellaterra (Barcelona), Spain}

\affiliation{D\'{e}partement de Physique Appliqu\'{e}e,  Universit\'{e} de Gen\`{e}ve,  1211 Gen\`{e}ve,  Switzerland}

\title{
Fundamental limits of metrology at thermal equilibrium
}

\begin{abstract}
    We consider the estimation of an unknown parameter $\theta$ through  a quantum probe at thermal equilibrium.  The probe is assumed to be in a Gibbs state  according to its Hamiltonian~$H_\theta$, which  is divided in 
    a parameter-encoding  term $\hp_\theta$   and an additional, parameter-independent, control~$\hc$. Given a fixed encoding, we find the maximal Quantum Fisher Information attainable via arbitrary~$\hc$, which provides a fundamental bound on the measurement precision. We elucidate the role of quantum coherence between encoding and control in different temperature regimes, which include ground state metrology as a limiting case.
    In the case of locally-encoded parameters, the optimal sensitivity presents a $N^2$-scaling in terms of the number of particles of the probe, which can be reached, at finite temperature, with local measurements and no entanglement. 
    We apply our results to paradigmatic spin chain models, showing that these fundamental limits can be approached using local two-body interactions.
    Our results set the fundamental limits and optimal control for metrology with thermal and ground state probes, including probes at the verge of criticality. 
\end{abstract}

\maketitle

\section{Introduction}
Quantum metrology studies how to precisely estimate an unknown physical parameter $\theta$  and the ultimate limits that quantum physics sets on the precision~\cite{paris2009quantum,giovannetti2011advances,Degen2017}. 
Typically, when modeling the estimation process, 
$\theta$ enters in the Hamiltonian of the probe system $H_\theta$.
In consequence, its state $\rho_\theta$ eventually correlates to the parameter value, which can be then estimated by measuring the probe. 
Through the Cr\'amer-Rao bound~\cite{cramer1999mathematical,rao1945information}
\begin{equation} \label{eq:CR_main}
\Delta \theta^2\geq \frac{1}{n \F_\theta}
\end{equation}
the minimum error attainable by any locally-unbiased estimator is  related to the quantum Fisher Information (QFI) $\F_\theta$ of the state $\rho_\theta$ and the number of repetitions of the experiment $n$. The QFI is thus a central quantity in metrology, which measures the sensitivity $\rho_\theta$  to small variations of $\theta$~\cite{helstrom_quantum_1969,holevo_probabilistic_1982,braunstein_statistical_1994}. 

A standard setting considered in the literature is that of \emph{dynamical} metrology. In this case, the probe is prepared in a well controlled initial state and is left to coherently evolve under the influence of the Hamiltonian $H_\theta$ for a time $t$. 
The QFI of the probe's final state is there bounded by the maximum variance of $(t/\hbar)\partial_\theta H_\theta$~\cite{boixo2007generalized,pang2014quantum} (cf. App.~\ref{app:dyna_compa}), which, for the paradigmatic $N$-partite case $H_\theta=\theta\sum_{k=1}^N h^{(k)}$, can scale at most as  
\begin{align}
\mathcal{F}_\theta  \lesssim \frac{N^2 t^2}{\hbar^2} \quad\text{(dynamical),} 
\label{eq:HeisenbergLimit}
\end{align} 
known as the Heisenberg limit~\cite{giovannetti_quantum_2006} (notation $A \lesssim B$ indicates the existence of a scaling-independent $c>0$ s.t. $A\leq c B$).

Another well studied setting consists in estimating~$\theta$  from a steady state $\rho_\theta$ of an open quantum system~\cite{Braun2018,montenegro2024quantum}.
This can be a non-equilibrium  steady state~\cite{Banchi2014,Macieszczak2016,Samuel2017Quantum,Marzolino2017Fisher,pavlov2023quantum}, but the most common setting is the one of \emph{equilibrium} (including \emph{ground state}) metrology, where the parameter is encoded in the Gibbs (resp. ground) state of the Hamiltonian $H_\theta$
\begin{align}
\label{eq:thermal_state}
    \rho_\theta= \frac{e^{-\beta H_\theta}}{\Tr{e^{-\beta H_\theta}}},
\end{align}
where $\beta=1/(k_B T)$, $T$ is the temperature of the surrounding environment and $k_B$ is Boltzmann constant. 
The potential of Gibbs (ground) states in quantum metrology has been long recognised in the literature, particularly close to thermal (quantum) phase transitions~\cite{zanardi2007bures,Zanardi2008}. 
Expressions for the QFI of Gibbs states  were derived in
~\cite{zanardi2007bures,jiang2014zhang}. The possibility  of a measurement precision beyond the shot-noise limit ($\F_\theta \lesssim N$)  has been reported for thermal/ground states of spin systems~\cite{Invernizzi2008,mehboudi2016achieving,Rams2018,Montenegro2021Global,Salvia2023Critical,garcíapintos2024estimation}, light-matter interacting systems~\cite{Bina2016Dicke,Garbe2020,Ying2022}, and  first experimental realisations are currently being developed~\cite{liu_experimental_2021,Ding2022}. 
 For temperature estimation~\cite{Mehboudi_2019}, the form of optimal Hamiltonian maximizing the QFI at a given $\beta$ is known~\cite{Correa2015,Abiuso2022Optimal}. 


Despite this remarkable progress, the fundamental limits of equilibrium metrology remain unsettled. The establishment of such saturable upper-bounds is the main goal and result of this letter, see Table \ref{tab:comp}. 
For that, we consider $H_\theta$ consisting of two 
terms 
\begin{align}
\label{eq:H_P+H_C}
    H_\theta =  \hp_\theta+\hc,
\end{align}
where $\hp_\theta$ describes the (fixed) physical mechanism imprinting $\theta$ on the probe, whereas $\hc$ is  under experimental control. 
We then optimize the QFI of the state in Eq.~\eqref{eq:thermal_state} over all possible $\hc$, thus finding a general upper bound on it. For a $N$-body probe on which $\hp_\theta$ acts locally, our results can be summarised as follows. 
At \emph{any} temperature, the QFI of the Gibbs state at best scales as  
\begin{align}\label{eq: intro general}
\mathcal{F}_{\theta} \lesssim  N^2 \beta^2 \quad\text{(thermal equilibrium)}.
\end{align}
Crucially, this bound can be attained without any entanglement between the $N$ subsystem, but only through classical correlations induced by $\hc$. 
At \emph{zero} temperature this bound diverges. However, for systems with a minimal gap $\Delta$ we find that the QFI scales, in the low temperature limit, $\beta^{-1}\ll \Delta$, at most as 
\begin{align}\label{eq: intro gapped}
\mathcal{F}_{\theta} \lesssim \frac{N^2}{\Delta^2}\quad\text{(ground state with gap)}.
\end{align} 
  This bound can also be saturated by a proper choice of~$\hc$. In this case quantum coherence in \eqref{eq:thermal_state} is crucial, and it yields an exponential  advantage in $\beta \Delta$  with respect to the ``classical" commuting case $[H_\theta,\partial_\theta H_\theta]=0$.

  These results~\eqref{eq: intro general}, \eqref{eq: intro gapped}
  provide fundamental bounds on equilibrium metrology, analogously to Heisenberg's limit in  dynamical metrology~\eqref{eq:HeisenbergLimit}.  As shown in Table~\ref{tab:comp}, they can be generalized to arbitrary encodings. 

\begin{table}[]
    \centering
    \begin{tabular}{c|c}
 \sf{Encoding} & \sf{Maximum} $\F_\theta$ 
 \\ \hline \rule{0pt}{4ex} 
  \text{Dynamical}  $\rho_\theta = e^{-i H_\theta t/\hbar}\rho  e^{i H_\theta t/\hbar}$ & \hspace{1mm} $
  \dfrac{t^2}{\hbar^2} \|\partial_\theta H_\theta\|^2$ 
  (Refs. \cite{pang2014quantum,boixo2007generalized}) \rule[-2.5ex]{0pt}{0pt}  \\ \hline \rule{0pt}{4ex}    
 \text{Thermal} $ \rho_\theta= \dfrac{e^{-\beta H_\theta}}{\Tr{e^{-\beta H_\theta}}} $ & \hspace{1mm}  $\dfrac{\beta^2}{4}  \|\partial_\theta H_\theta\|^2$ \hspace{1.7mm} 
 (Eq. \eqref{eq:maxFI}) \rule[-2.5ex]{0pt}{0pt}\\ \hline \rule{0pt}{4ex} 
 Ground state  
 of $H_\theta$ \\ with spectral gap $\Delta$ & \hspace{1.7mm}  $\dfrac{1}{\Delta^2} \|\partial_\theta H_\theta\|^2$  \hspace{1mm} (Eq. \eqref{eq:maxQFI_gap})   \rule[-2.5ex]{0pt}{0pt} \\ \hline
\end{tabular}
    \caption{{\bf Main results.} 
    Given $H_\theta = \hp_\theta + \hc$ with fixed encoding $\hp_\theta$, we find the maximum of $\F_\theta$ for thermal and ground states of $H_\theta$ (by maximizing over all $\hc$). The maxima share close resemblance with the one for the dynamical encoding~\cite{boixo2007generalized}.  All of them depend on the seminorm $\|\partial_\theta H_\theta\|= {\lambda}_{\rm M}-{\lambda}_{\rm m}$, where $\lambda_{\rm M}$ ($\lambda_{\rm m}$) is the maximum (minimum) eigenvalue of $\partial_\theta H_\theta$ ($H'$ in the text). For a  $N$-partite probe $H_\theta=\sum_{i=1}^N h^{(k)}_\theta+\hc$ with bounded $\|\partial_\theta h^{(k)}_\theta\|$, we recover the scalings in  Eqs.~\eqref{eq:HeisenbergLimit}, \eqref{eq: intro general} and ~\eqref{eq: intro gapped}.}   
    \label{tab:comp}
\end{table}

\section{Estimation via measuring equilibrium states} 
 
  {To derive such results, it will become useful to introduce the Bures multiplication superoperator~\cite{scandi2023quantum} 
\begin{align} \label{eq: SLD}
    \J_{\rm B,\rho}[A]:=\frac{1}{2} (\rho A+A\rho)\;,
\end{align}
which has inverse $
 \J_{{\rm B},\rho}^{-1}[A]=2\int_0^\infty\D s \; e^{-\rho s} A e^{-\rho s}$ on positive full rank states $\rho$. 
 The operator $\J^{-1}_{\rm B,\rho_\theta}[\partial_\theta{\rho}_\theta]$ is the so-called \emph{symmetric-logarithmic-derivative} (SLD) of $\rho_\theta$, which can be used to generically express the QFI as~\cite{paris2009quantum}
 \begin{align}
\label{eq:QF_def}
\F_\theta:=\Tr{\partial_\theta{\rho}_\theta \,\J^{-1}_{\rm B,\rho_\theta}[\partial_\theta{\rho}_\theta]}\;.
\end{align}
}
 
Let us now apply \eqref{eq:QF_def} to  generic systems at thermal equilibrium represented by the Gibbs state $\rho_\theta$~\eqref{eq:thermal_state} at given temperature $T=(k_B \beta)^{-1}$.
In order to compute the QFI~\eqref{eq:QF_def} one can express the variation of $\rho_\theta$~\eqref{eq:thermal_state} via the operator exponential derivative~\cite{hiai2014introduction}
as
 {
$\partial_\theta{\rho}_\theta=-\J_{\rm L,\rho_\theta}[\beta \partial_\theta{H}_\theta]+\rho_\theta\Tr{\rho_\theta\beta \partial_\theta H_\theta}$},
where we introduced the logarithmic multiplication superoperator~\cite{scandi2023quantum}
\begin{align}\label{eq: new product}
    \J_{\rm L,\rho}[A]:=\int_0^1 \D s\; \rho^s A \rho^{1-s}\;.
\end{align}
 {In what follows, we will denote $\rho:=\rho_\theta$ for simplicity as well as $H':=\partial_\theta H_\theta\equiv\partial_\theta \hp_\theta$.}
By substituting  the above expressions in Eq.~\eqref{eq:QF_def} and simplifying the cross-terms (see Appendix for details) we find {the general expression of $\F_\theta$ for perturbations of systems at thermal equilibrium} (see also~\cite{zanardi2007bures,jiang2014zhang,scandi2023quantum}), that is
\begin{align}
\label{eq:QFI_J}
    \F_\theta 
    =\beta^2\left(\Tr{H'\mathcal{J}_{\rho}[H'] } - \Tr{\rho H'}^2\right)\;,
\end{align}
corresponding to the generalized variance of $H'$,
according to the multiplication superoperator  {defined by the composition}
$
    \mathcal{J}_\rho:=\J_{\rm L,\rho}\circ \J^{-1}_{\rm B,\rho}\circ\J_{\rm L,\rho}\;.
$
Notice that $\J_{\rm B,\rho},\J_{\rm L,\rho},\mathcal{J}_\rho$ can all be analytically expressed in the operator basis $\ketbra{i}{j}$ using the eigenvectors of $\rho$ (cf.~\cite{scandi2023quantum}).

As a basic example, 
consider now a single qubit,   {with} an encoding  {$\hp_\theta=\theta\sigma_z$} and control  {(up to rotational symmetry)} $\hc= \Delta
\left( \cos(\alpha)\sigma_z +\sin(\alpha) \sigma_x\right)$ in~\eqref{eq:H_P+H_C}. Using~\eqref{eq:QFI_J} one finds  {at~$\theta=0$}
\begin{equation}
    {\F_{\theta=0}^{\rm qubit}} = \beta ^2\left(\frac{\cos ^2(\alpha )}{\cosh ^2(\beta  \Delta )}+\frac{\sin ^2(\alpha ) \tanh ^2(\beta  \Delta )}{(\beta  \Delta )^2}\right).
\end{equation}
Without further constraints, the optimal control is then obtained by choosing $\hc$ to be degenerate $(\Delta=0)$, for which the QFI is maximized to $\beta^2$. 
Instead, assuming a minimum energy gap $\Delta$, the QFI becomes dependent on the angle $\alpha$ between $H'$ and $\hc$. If encoding and control commute $(\alpha=0)$ the QFI decays exponentially in the low-temperature limit $\beta\Delta\gg 1$,  {$\F_{\theta=0}^{\rm qubit} 
\to 4 \beta^2 e^{-2 \beta \Delta}$}. In contrast, for maximally non-commuting $H'$ and $H_C$ $(\alpha=\frac{\pi}{2}$), the QFI remains finite and is maximal in the same limit,  {$\F_{\theta=0}^{\rm qubit} 
\to  1/\Delta ^2$}.
Remarkably, we will find that these properties are reflected in the maximization of the QFI of thermal probes of arbitrary dimension.

\section{Ultimate bounds for the Fisher Information at thermal equilibrium}
Starting from the generic QFI for a system at equilibrium~\eqref{eq:QFI_J}, we now find the choice of controls $\hc$ that maximizes it.
{For that,} first consider the following bound valid for any $\rho$ and $H'$
\begin{equation}
\label{eq:main_ineq}
\Tr{H'\mathcal{J}_{\rho}[H']} - \Tr{\rho H'}^2 
\leq \text{Var}_{\rho}[H'].
\end{equation}
This is a direct consequence of the superoperator inequality $\mathcal{J}_\rho\leq\J_{\rm B,\rho}$ and is saturated for $[\rho,H']=0$, see~\cite{scandi2023quantum,petzghinea_introduction}. Second, note that the variance $\text{Var}_{\rho}[H'] := \Tr{\rho H'^2 } - \Tr{\rho H'}^2$ is upper bounded by 
$({\lambda}_{\rm M}-{\lambda}_{\rm m})^2/4$, where $\lambda_{\rm M(m)}$ is the maximum(minimum) eigenvalue of $H'$. This standard bound, derived for completeness in the Appendix~\ref{app:std_var_ineq}, is tight for $\rho=\frac{1}{2}({\ketbra{\lambda_{\rm M}}{\lambda_{\rm M}}+\ketbra{\lambda_{\rm m}}{\lambda_{\rm m}}})$ which commutes with~$H'$. Such state can be well approximated by the Gibbs state of the Hamiltonian $\hp_\theta+\hc=\epsilon (\ketbra{\lambda_{\rm M}}{\lambda_{\rm M}}+\ketbra{\lambda_{\rm m}}{\lambda_{\rm m}})+ H_{\perp}$ where $H_{\perp}$ is orthogonal to ${\rm Span}(\ket{\lambda_{\rm M}},\ket{\lambda_{\rm m}})$ and has all its eigenvalues $\eta_{\perp j}$ satisfying $\beta(\eta_{\perp j}-\epsilon)\gg 1$, leading to exponentially small corrections in the approximation.
 Connecting  {these observations to the inequality~\eqref{eq:main_ineq}} we obtain the fundamental bound  {for the thermal QFI~\eqref{eq:QFI_J}}
\begin{align}
     {\F_\theta\leq \beta^2 \frac{\|H'\|^2}{4}\;,}
    \label{eq:maxFI}
\end{align}
 {where we have defined the operator seminorm: $\|H'\|= {\lambda}_{\rm M}-{\lambda}_{\rm m}$~\cite{boixo2007generalized}. }
This is \emph{the ultimate upper bound to the QFI at finite temperature}, for a given encoding of $\theta$ that locally behaves as $H'$.
Moreover, as~\eqref{eq:maxFI} is saturated for commuting $[\rho,H']=0$  {(equivalently $[\hp_\theta+\hc,H']=0$)} we see that  {\emph{no fundamental advantage is provided by quantum coherence} in (thermal) equilibrium metrology}. Notice that this is valid assuming full control on $\hc$ and finite equilibrium temperature. The $\beta^2$ factor makes this bound trivial in the limit of zero temperature. Below we discuss this limit,  {in which we recover a coherence-advantage considering physical constraints to the control.}

A striking immediate consequence of the bound~\eqref{eq:maxFI} can be obtained when considering the case of probes composed of $N$ subsystems. When the parameter is encoded locally, i.e. $H'=\sum_{k=1}^{N} h^{(k)}$ with the operator $ h^{(k)}$ acting on the $k$-th system and with  $\|h^{(k)}\| \leq  \|h\|$ , a direct application of the bound gives 
\begin{align}
\label{eq:F_hei_scal}
   \F_\theta^\text{local}\leq \beta^2 \frac{\|h\|^2 N^2}{4}\;. 
\end{align} 
Hence, \emph{at finite temperature, the QFI relative to a local parameter scales at most as $N^2$ in the system's size}.
The quadratic $N^2$ scaling reminds of the well-known Heisenberg limit~\eqref{eq:HeisenbergLimit} of quantum metrology~\cite{giovannetti_quantum_2006,giovannetti2011advances}, however  our bounds are saturated for classically correlated states. 
In particular when $H'$ is a local perturbation, the optimal preparation $\propto e^{-\beta(\hp_\theta+ \hc)}$ \emph{does not feature entanglement}, and the measurement basis can be chosen to be local.


Here 
we recall  that the quantum Cramér-Rao bound~\eqref{eq:CR_main} is saturated by choosing to measure in the basis of the SLD~\cite{paris2009quantum}, which in our notation is simply given by $\J^{-1}_{\rm B,\rho_\theta}[\partial_\theta\rho_\theta]$~\cite{paris2009quantum}. In the commuting case $[\hp_\theta,H']=0$ this optimal measurement basis clearly coincides with any choice diagonalizing $\hp_\theta, H',\rho$ (also valid under milder assumptions~\cite{salmon2023onlyclassical}), while in general it might not coincide with any of the former. In the Appendix~\ref{app:SLD_and_meas} we analyze the performance of energy measurements in different basis. 


\section{Low temperature limit and quantum advantage}
The previous discussion shows that at finite temperature the  {optimally-controlled} Hamiltonian  {$H_\theta=\hp_\theta+\hc$} commutes with $H'$, and can ensure a maximum sensitivity~\eqref{eq:maxFI} that diverges in the limit of small temperature.
However, in order to saturate~\eqref{eq:maxFI} it needs the max/min eigenstates $\ket{\lambda_{\rm M/m}}$ of $H'$ for (doubly degenerate) ground states.
Clearly, the implementation of such controlled Hamiltonian is in general non-trivial. 
Let us now assess the robustness of the ultimate precision limit when subjected to control restrictions. 
In particular, we now consider the QFI maximisation while constraining $H_\theta$ 
to have a unique ground state with a minimum energy gap $\Delta$ to the first excited state.


To tackle this case it is useful to express the QFI~\eqref{eq:QFI_J} in the eigenbasis of {$H_\theta$}, that is $H_\theta\ket{i}=E_i\ket{i}$, which diagonalises the unperturbed thermal state as $\rho=\sum_i p_i \ketbra{i}{i}$\; with $p_i =e^{-\beta E_i}/(\sum_j e^{-\beta E_i})$. As shown in App.~\ref{subsec:CQ_split} (see also \cite{Zanardi2008}), 
\begin{align}
\F_\theta 
= \F_\theta^{\rm diag} + \sum_{i,j|E_i < E_j} \frac{4(p_i-p_j)^2|H'_{ij}|^2}{(E_i - E_j)^2(p_i+p_j)} \;.
\label{eq:QFI_components}
\end{align}
Here 
$\F_\theta^{\rm diag}= \beta^2 \,\text{Var}_{\rho}[\sum_E \Pi_E H'\Pi_E]$ corresponds to the classical variance of the dephased $H'$, as $\Pi_E =\sum_{i|E_i=E} \ketbra{i}$ is the projector on the subspace of energy $E$. 
Notice that for classical 
configurations $[H_\theta,H']=0$ clearly $\F_\theta=\F_\theta^{\rm diag}$.


Imposing a non-degenerate ground state $\ket{0}$ with a minimum energy gap $E_i-E_0\geq \Delta$ to all other levels implies that in the low temperature limit the populations of the excited levels ($i>0$) decay exponentially $p_{i} =\mathcal{O}(e^{-\beta \Delta })$. In consequence, the classical contribution to the QFI is also suppressed exponentially.
The situation is markedly different for the off-diagonal contribution in~\eqref{eq:QFI_components}.
While the terms with $i>0$ decay exponentially, the terms with $i=0$ remain constant as $\frac{(p_0-p_j)^2}{p_0+p_j}\to 1$. In the large gap/low temperature limit the dominant contribution thus becomes
\begin{align}
   \F^\text{low-T}_{\theta}= \sum_{i>0} \frac{4|H'_{0i}|^2}{(E_i-E_0)^2} 
    + \mathcal{O}(e^{-\beta\Delta})\;.
    \label{eq:QF_lowtemp}
\end{align}
 We see immediately that  {\emph{coherence provides an exponential advantage at low temperature} (or large gap limit)}. We recognize~\eqref{eq:QF_lowtemp} as the norm of the ground state variation at first order perturbation theory. This is expected in the low temperature limit where $\rho_\theta$ collapses to the ground state of $H_\theta$. 
 
 In order to maximize~\eqref{eq:QF_lowtemp} w.r.t. all possible $\Delta$-gapped $H_\theta$, consider the following inequality 
\begin{align}
    \sum_{i>0} \frac{|H'_{0i}|^2}{(E_i-E_0)^2} \leq \frac{\sum_{i>0}|H'_{0i}|^2}{(E_1-E_0)^2}=
    \frac{{\rm Var}_{\ket{0}}[H']}{(E_1-E_0)^2}\;.
\end{align}
The inequality is saturated whenever $H'$ only connects the ground state $\ket{0}$ to the first excited level. Moreover, as we are free to choose the ground state $\ket{0}$ of $H_\theta$, this should be chosen in order to maximise the variance of $H'$, which is maximal for states satisfying $|\braket{\lambda_{\rm M}}{0}|^2=|\braket{\lambda_{\rm m}}{0}|^2=\frac{1}{2}$. 
These considerations, together with the gap assumption $E_1\geq \Delta$, lead us to the bound
\begin{align}
     \F^\text{low-T}_\theta \leq \frac{\|H'\|^2}{\Delta^2}+\mathcal{O}(e^{-\beta\Delta})\;,
\label{eq:maxQFI_gap}
\end{align}
which is saturated e.g. by choosing $\ket{0}=\frac{\ket{\lambda_{\rm M}}\pm\ket{\lambda_{\rm m}}}{\sqrt{2}}$ and $\ket{1}=\frac{\ket{\lambda_{\rm M}}\mp\ket{\lambda_{\rm m}}}{\sqrt{2}}$. 
Eq.~\eqref{eq:maxQFI_gap} sets the ultimate limit of estimation in low-temperature gapped ground states. We notice that, while for finite temperature the scale of the QFI~\eqref{eq:maxFI} is dictated by the perturbation's variance in thermal energy units $\beta^2{\rm Var}[H']$, here the scaling is dictated rather by the ratio with the gap, ${\rm Var}[H']/\Delta^2$. As a direct consequence of~\eqref{eq:maxQFI_gap}, similarly to Eq.~\eqref{eq:F_hei_scal}, the QFI relative to a local parameter cannot scale faster than $N^2$, \emph{unless the gap $\Delta$ decreases in the system size $N$}~\cite{Rams2018}.


 {
Let us now relate our bound with previous metrology results  with ground state probes prepared close to a quantum phase transition~\cite{Invernizzi2008,Rams2018,montenegro2024quantum}. There,  the QFI scales as $\F_\theta \sim N^{2/d\nu}$, where $d$ is the physical dimension and $\nu$ a critical exponent~\cite{Albuquerque2010,montenegro2024quantum}. From our bound, it follows that necessarily  $N^{2/d\nu} \lesssim  N^2/\Delta^2  $. 
When including the spectral gap critical scaling $\Delta \sim N^{-z/d}$, 
this immediately yields the exponent inequality: 
    $\nu (d+z) \geq 1$, 
 which is consistent with established results~\cite{continentino1994quantum,cardy1996scaling}. 
Moreover, 
our results not only provide a scaling inequality, but an exact bound on $\F_\theta$ that is applicable to arbitrary  Gibbs states (at criticality or away from it). 
}

\section{High temperature limit} 
For completeness, we report  the opposite limit of high temperature $\beta\rightarrow 0$. In such case we can approximate the state by $\rho_\theta\approx \frac{\eye-\beta H_\theta}{\Tr{\eye-\beta H_\theta}}\approx \frac{1}{D}(\eye-\beta \tilde{H}_\theta)$, where $D$ is the dimension of the Hilbert space and $\Tilde{H}_\theta:=H_\theta-\Tr{H_\theta}$ is the trace-less Hamiltonian. Interestingly, one can prove (App.~\ref{app:J_loc_eye}) that close to maximally mixed states $\J_{{\rm B},\eye+\varepsilon}\approx \J_{{\rm L},\eye+\varepsilon}\approx\mathcal{J}_{\eye+\varepsilon}$ all coincide up to $\mathcal{O}(\varepsilon^2)$ corrections. It follows that
~\eqref{eq:QFI_J} simplifies for high temperatures as
\begin{align}
 \F^\text{high-T}_\theta=\beta^2\left(\Tr{H'^2\rho}-\Tr{\rho H'}^2\right) + \mathcal{O}(\beta^4)\;.
\end{align}
More precisely, the  {leading} order $\beta^2$ in the above equation is given by approximating $\rho=\eye/D$  {and is thus control-independent}, while the $\mathcal{O}(\beta^3)$ term  {can be computed as $-D^{-1}\beta^3\Tr{\tilde{H}_{\theta}\tilde{H'}^2}$.}

\section{Example. 1D Ising Chain}
\label{sec:examples}

Let us now discuss how to approach these fundamental limits with the paradigmatic  1D Ising model
\begin{align}\label{eq: ID ising H}
    H_\theta = - J \sum_{i=1}^N Z_i Z_{i+1} + (B+\theta_Z) \sum_{i=1}^N Z_i + \theta_X \sum_{i=1}^N X_i 
\end{align}
with periodic boundary conditions. 
Here  $Z_i=\sigma_z^{(i)}$ and $X_i = \sigma_x^{(i)}$ are Pauli operators acting on the $i$-th spin. 

\begin{figure}
    \centering
\includegraphics[width=\columnwidth]
    {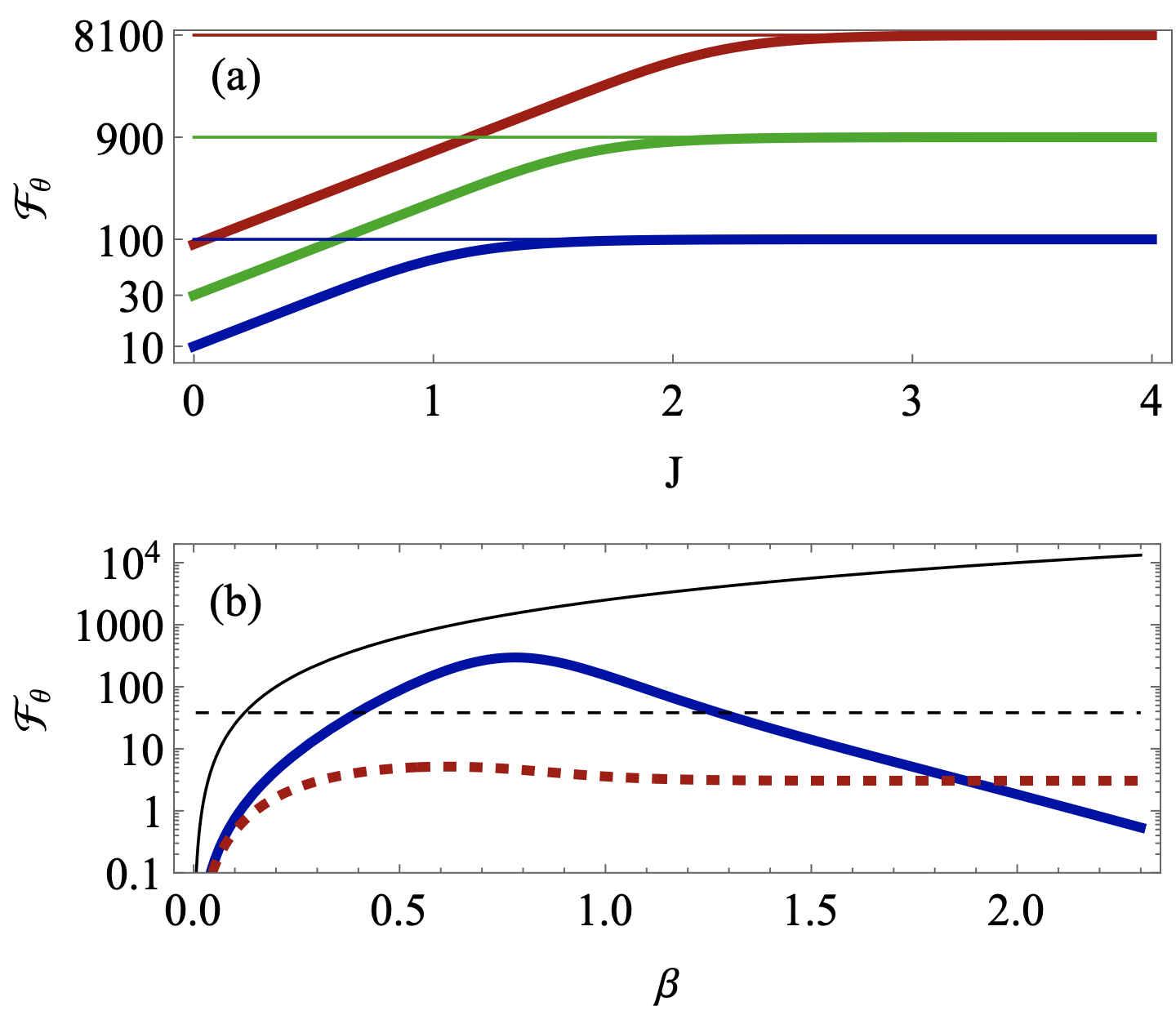}
    \caption{ The Quantum Fisher information of the Gibbs state of the 1D Ising model with respect to a parallel field $\theta_Z$ (full lines) and transverse field $\theta_X$ (dashed line in (b)).  The thin horizontal lines depict the upper-bound $\F_\theta\leq \beta^2 N^2$ in Eq.~\eqref{eq:maxFI}.  \textbf{(a)} As function of $J$ for $B=0$. Here $\beta=1$ and the number of spins is from bottom to top  $N=10$ (blue), $30$ (green) and $90$ (red). \textbf{(b)} As function of the parameter $\beta$. Here the number of spins is $N=50$, $B=0.05$ and $J=2$. The thin dashed line depicts the  upper-bound $\F_\theta\leq\frac{N^2}{4(B+2J)^2}$ in the low temperature limit of Eq.~\eqref{eq:maxQFI_gap}. }
    \label{fig:ising_strength}
\end{figure}

We first consider the estimation of a parallel field $\theta_Z$  (at  $\theta_Z=\theta_X=0$). 
As shown earlier, the QFI for this classical model is the variance $\F_{\theta_Z=0} =\beta^2 \, \text{Var}_{\rho}[\sum_i Z_i]$. This can be computed analytically from the partition function, as detailed in  Appendix~\ref{app: Ising}. In particular, if $B=0$ and $\beta J$ increases faster than $\ln N$, we find that the thermal QFI reaches the upper limit~\eqref{eq:F_hei_scal} $\F_{\theta_Z=0} = \beta^2 N^2 + \mathcal{O}( N^4 e^{- 4 \beta J })$ (see also Fig.~\ref{fig:ising_strength}). This can be understood from the observation that in such a strong ferromagnetic regime the Gibbs state  satisfies $\rho \approx \frac{1}{2}(\ketbra{\uparrow}^{\otimes N} + \ketbra{\downarrow}^{\otimes N})$ as desired. 
However, when $B>0$,  the QFI decays exponentially in $\beta B$, due to the energy splitting between $\ketbra{\uparrow}^{\otimes N}$ and $\ketbra{\downarrow}^{\otimes N}$.

Next, we consider the estimation of the transverse field $\theta_X$ (at $\theta_Z=\theta_X=0$).  In this case the QFI can be expressed through the correlations of three neighbouring spins $\F_{\theta_X=0} = \sum_i \, \Tr{\rho \, \zeta(Z_{i-1},Z_i,Z_{i+1})}$ and computed analytically, as detailed in the Appendix. 
It is insightful to consider the low temperature limit. For $B > 0$, the unique ground state $\ket{0}= \ket{\downarrow}^{\otimes N}$ is coupled to a single eigenstate $\ket{1_W} = \frac{1}{\sqrt N} \sum_{i} X_i \ket{0}$ of the unperturbed Hamiltonian, which has the energy $E_1 = E_0 + 2 B + 4 J$. Hence, by Eq.~\eqref{eq:QF_lowtemp} we find the leading order contribution to the QFI to be $ \mathcal{F}_{\theta_X=0} = \frac{4 |\bra{0} \sum_{i} X_i \ket{1_W}|^2}{(E_1-E_0)^2}= \frac{N}{ (B+2J)^2}$,
which remains constant at $\beta\to\infty$. This  is a factor $N/4$ smaller compared to the upper bound \eqref{eq:maxQFI_gap}. Instead, as pointed out before, a classical strategy would decay exponentially with~$\beta$.  These results are illustrated in Fig.~\ref{fig:ising_strength}. 

\section{Comments and conclusion}




In this work we have derived fundamental bounds on equilibrium 
quantum metrology, where a parameter $\theta$ is encoded in a system at thermal equilibrium
~\eqref{eq:thermal_state}. Assuming full control on the parameter-independent part of the system's Hamiltonian, we derived the upper limit~\eqref{eq:maxFI} to the QFI.
This upper-bound is shown to be attained by a ``classical''  {commuting strategy in which} $[H_\theta,H']=0$. In the low temperature limit the bound~\eqref{eq:maxFI} diverges.
This motivated us to strengthen it to Eq.~\eqref{eq:maxQFI_gap}, which accounts for the presence of a spectral gap $\Delta>0$  and remains finite when $\beta \to \infty$. To saturate the low-temperature bound~\eqref{eq:maxQFI_gap},  quantum coherence is crucial and in fact an exponential gap appears in the QFI  {between non-commuting ($[H_\theta,H']\neq 0$) and commuting strategies}. When $\theta$ is encoded locally on a $N$-body probe, \eqref{eq:maxFI} and~\eqref{eq:maxQFI_gap} display a   quadratic scaling QFI $\propto N^2$. Finally we showcased our results on paradigmatic classical and quantum spin chains, in particular showing that a  1D classical spin chain probe in the strongly interacting ferromagnetic regime 
can approach the fundamental limit~\eqref{eq:maxFI}.

Our first main result~\eqref{eq:maxFI} opens a clear avenue for the design of optimal thermal probes for equilibrium metrology beyond the case of thermometry~\cite{Correa2015,Mehboudi_2019,Campbell_2018Precision,Mok2021,Abiuso2022Optimal,Brenes2023Multispin}. 
{Compared to recent works providing bounds for parameter estimation on thermal states~\cite{garcíapintos2024estimation}, or studying specific models, we crucially \emph{solve} the control-maximisation of QFI, showing that beyond shot-noise error is possible~\cite{Invernizzi2008,mehboudi2016achieving,Salvia2023Critical,garcíapintos2024estimation}, but is ultimately limited, at finite temperature, by a quadratic scaling of the QFI, which is achievable remarkably by separable states.}
Likewise, our second main result~\eqref{eq:maxQFI_gap} is helpful to understand the limits of ground state metrology~\cite{Zanardi2008,mohammdi2023metrology}, 
 {including the case of probes at quantum phase transitions and their critical scaling~\cite{Invernizzi2008,Rams2018,montenegro2024quantum,Albuquerque2010}.}
bound~\eqref{eq:HeisenbergLimit}.  

{Our approach may provide a new avenue to investigate tradeoffs between thermalization time and enhancements in measurement precision~\cite{Anto-Sztrikacs2023}, and to characterize fundamental thermalization timescales~\cite{Hartnoll2022}.}
Finally, our results might provide insights to  the study of Hamiltonian learning, which has been showed to saturate the Heisenberg limit in the dynamical setting~\cite{huang2023learning,dutkiewicz2023advantage}, and is being intensively studied in the thermal scenario~\cite{anshu2021sample-efficient,haah2022optimal,bakshi2023learning}.

\section*{Acknowledgements}

The authors thank Ricard Puig i Valls and Paolo Andrea Erdman for valuable discussions.

\begin{wrapfigure}{r}{2cm}
\includegraphics[width=2cm]{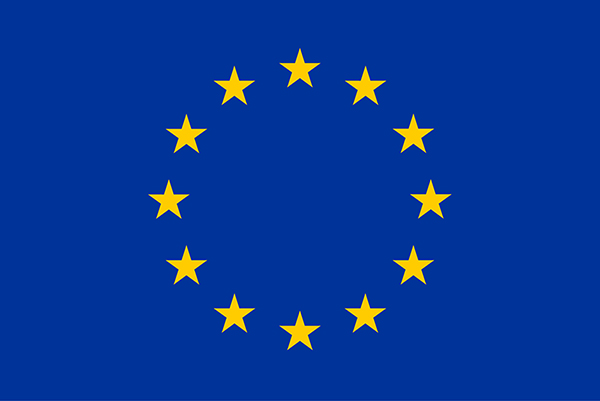}
\end{wrapfigure}
P.A. is supported by the 
QuantERA II programme, that has received funding from the European Union’s Horizon 2020 research and innovation programme under Grant Agreement No 101017733, and from the Austrian Science Fund (FWF), project I-6004. P.S. is supported by the NCCR SwissMAP.
J.C.  is supported by the QuantERA grant C’MON-QSENS!, by Spanish MICINN PCI2019-111869-2,  AEI PID2022-141283NB-I00, and MCIN with funding from European Union NextGenerationEU (PRTR-C17.I1) and by Generalitat de Catalunya. J.C. also acknowledges  support from ICREA Academia award. M. P.-L. acknowledges support from the Spanish Agencia Estatal de Investigacion through the grant  ``Ram{\'o}n y Cajal RYC2022-036958-I''  and the Swiss National Science Foundation through an Ambizione Grant No. PZ00P2-186067.

\bibliography{equilibriumBIB}

\newpage
\
\newpage

\widetext

\appendix

\section{Analytical expressions for the Quantum Fisher Information}
\label{app:analytical}

In this section, we give additional details for the main expressions used in the main text.
We remind first the definitions of the superoperators we used,
\begin{align}
\label{eqapp:J_bur}
    \J_{\rm B,\rho}[A] &:=\frac{\rho A+A\rho}{2}\;, &   \J^{-1}_{\rm B,\rho}[A] &= 2\int_0^\infty\D s \; e^{-\rho s} A e^{-\rho s}\;, \\
\label{eqapp:J_log}
    \J_{\rm L,\rho}[A] &:=\int_0^1 \D s\;\rho^{s}A\rho^{1-s}\;, &   \J^{-1}_{\rm L,\rho}[A] &= \int^{\infty}_0 \D s \frac{1}{\rho+s}A\frac{1}{\rho+s}\;, \\
\label{eqapp:J_fish}
    \mathcal{J}_{\rho}[A] &:=\J_{\rm L,\rho}\circ \J^{-1}_{\rm B,\rho}\circ \J_{\rm L,\rho}[A]\;. & &  
\end{align}
Throughout this work, we consider $\rho$ to be positive and full rank. The results for pure states can be recovered by the appropriate limit $\rho\rightarrow \ketbra{\psi}{\psi}$.
The $\J$ operators here listed (including $\mathcal{J}$)
belong to the family of generalized quantum Petz-Fisher multiplications~\cite{petzMonotoneMetricsMatrix1996,scandi2023quantum} ($\mathcal{J}_\rho$ corresponds to the quantum information variance superoperator $\J_{\rm V,\rho}$ in~\cite{scandi2023quantum}), and represent non-commutative versions of the multiplication times $\rho$.
These operators are all self-adjoint w.r.t. the Hilbert-Schmidt scalar product, meaning that $\Tr{A\J[B]}=\Tr{\J[A]B}$ for Hermitian $A,B$. Notice also
and when fixing the base point $\rho$, they commute, i.e. for example $\J^{-1}_{\rm B,\rho}\circ\J_{\rm L,\rho}=\J_{\rm L,\rho}\circ \J^{-1}_{\rm B,\rho}$.
In addition to their operator-integral expressions above, notice that the action of these superoperators can be explicited in the components $\ketbra{i}{j}$, where $\ket{i}$ are the eigenvectors of $\rho$. That is, given
\begin{align}
    \rho=\sum_i p_i \ketbra{i}{i}\;,
\end{align}
one has
\begin{align}
\label{eqapp:J_B_coordi}
    \J_{\rm B,\rho}[\ketbra{i}{j}] &=\frac{p_i+p_j}{2} \ketbra{i}{j}\;,\\
\label{eqapp:J_L_coordi}
    \J_{\rm L,\rho}[\ketbra{i}{j}] &=\frac{p_i-p_j}{\ln p_i - \ln p_j} \ketbra{i}{j}\;.
\end{align}
It is important to notice that, while seemingly singular for $p_i=p_j$, the operator $\J_{\rm L,\rho}$ presents no singularities as $\J_{\rm L,\rho}[\ketbra{i}{i}]=p_i \ketbra{i}{i}$. This can be verified either by direct inspection of~\eqref{eqapp:J_log}, or by simply taking the limit
\begin{align}
\lim_{p_i\rightarrow p_j}    \frac{p_i-p_j}{\ln p_i - \ln p_j}=p_j\;.
\end{align}
It follows that $\J_{\rm L,\rho}[\ketbra{i}{i}]=\J_{\rm B,\rho}[\ketbra{i}{i}]=\rho \ketbra{i}{i}=p_i \ketbra{i}{i}$. More in general, we see that when restricting to commuting operators, they all simplify to a multiplication
\begin{align}
    [\rho,A]=0\quad \Rightarrow\quad \J_{\rm B,\rho}[A]=\J_{\rm L,\rho}[A]=\mathcal{J}_{\rho}[A]\equiv \rho A\;.
\end{align}

\subsection{Derivation of general QFI expression~\eqref{eq:QFI_J}}
In the following, we use dot notation $\dot{A}\equiv\partial_\theta A$ to identify variations with respect to $\theta$, in order to streamline the technical derivations.

The QFI relative to $\theta$ encoded in a system $\rho_\theta$ 
can be written as~\cite{paris2009quantum}
\begin{align}
\F_\theta:=\Tr{\dot{\rho}_\theta \J^{-1}_{\rm B,\rho_\theta}[\dot{\rho}_\theta]}\;.
\label{eqapp:Fishbur}
\end{align}
In the case we consider there $\rho_{\theta}=e^{-\beta H_\theta}/\Tr{e^{-\beta H_\theta}}$ is a Gibbs state and its variation follows from the exponential derivative~\cite{hiai2014introduction}
\begin{align}
   \partial_\theta e^{X_\theta}=\int_0^1 \D s\; e^{sX_\theta} \partial_\theta X_\theta e^{(1-s)X_\theta}\equiv \J_{\rm L,e^{X_\theta}}[\partial_\theta X_\theta]\;,
\end{align}
which can be applied immediately to see
\begin{align}
    \dot{\rho}_\theta= -\beta \frac{\J_{\rm L,e^{-\beta H_\theta}}[\dot{H}_\theta]}{\Tr{e^{-\beta H_\theta}}} + \beta \frac{e^{-\beta H_\theta}}{\Tr{e^{-\beta H_\theta}}^2}\Tr{\J_{\rm L,e^{-\beta H_\theta}}[\dot{H}_\theta]}=-\J_{\rm L,\rho_\theta}[\beta \dot{H}_\theta]+\rho_\theta\Tr{\beta \dot{H}_\theta\rho_\theta}\;,
\label{eqapp:rho_dot}
\end{align}
where we have used the properties $\J_{\rm L,\alpha\rho}=\alpha\J_{\rm L,\rho}$ for any scalar number $\alpha$, and $\Tr{\J_{\rm L,\rho}[A]}=\Tr{\rho A}$.

Once the expression~\eqref{eqapp:Fishbur} for $\dot{\rho}_\theta$ is given, it is a matter of substituting in~\eqref{eqapp:Fishbur} to obtain
\begin{align}
    \beta^{-2}\F_\theta=\Tr{\J_{\rm L,\rho_\theta}[\dot{H}_\theta] \J_{\rm B,\rho_\theta}^{-1} \circ \J_{\rm L,\rho_\theta}[\dot{H}_\theta] } - 2 \Tr{\J_{\rm L,\rho_\theta}[\dot{H}_\theta]\J^{-1}_{\rm B,\rho_\theta}[\rho_\theta]}\Tr{\rho_\theta \dot{H}_\theta}+ \Tr{\rho_\theta\J^{-1}_{\rm B,\rho}[\rho_\theta]}\Tr{\rho_\theta \dot{H}_\theta}^2\;.
    \label{eqapp:fish_with_cross}
\end{align}
The last terms of~\eqref{eqapp:fish_with_cross} can be simplified by noticing that $\J^{-1}_{\rho}[\rho]=\eye$ and $\Tr{\J_{\rho}[A]}=\Tr{\rho A}$ (these properties hold for $\J_{\rm L}$ as well as $\J_{\rm B}$, and all generalized Petz multiplication superoperators~\cite{scandi2023quantum}). Using these simplifications we arrive to~\eqref{eq:QFI_J}, that is
\begin{align}
 \beta^{-2}\F_\theta  = \Tr{\J_{\rm L,\rho_\theta}[\dot{H}_\theta] \J_{\rm B,\rho_\theta}^{-1} \circ \J_{\rm L,\rho_\theta}[\dot{H}_\theta] } - \Tr{\rho_\theta \dot{H}_\theta}^2\;.
 \label{eqapp:b-2_Fish}
\end{align}
Notice that $\J_{\rm L,\rho}$ and $\J_{\rm B,\rho}$ commute, and are self-adjoint with respect to the Hilbert-Schmidt scalar product, i.e. $\Tr{\J[A]B}=\Tr{A\J[B]}$ for hermitian $A,B$. This allows to express the QFI as in~\eqref{eq:QFI_J}, that is $\Tr{\dot{H}_\theta \mathcal{J}_{\rho_\theta}[\dot{H}_\theta] } - \Tr{\rho_\theta \dot{H}_\theta}^2$, with
\begin{align}
    \mathcal{J}_{\rho_\theta}=  \J_{\rm L,\rho_\theta}\circ \J_{\rm B,\rho_\theta}^{-1} \circ \J_{\rm L,\rho_\theta}
    \equiv \J_{\rm B,\rho_\theta}^{-1} \circ \J^2_{\rm L,\rho_\theta}\;.
\end{align}

\subsection{ {Locality of the Fisher Information and encoding-control decomposition}}
\label{sec:local_control_rationale}

In this section we comment the rationale beyond the physical choice of control scenario~\eqref{eq:H_P+H_C} for thermal states.
Equation~\eqref{eqapp:b-2_Fish} expresses in full generality the QFI for any differentiable encoding of $\theta$ in the Hamiltonian of the system, for thermal states at temperature $(k_{\rm B}\beta)^{-1}$. By direct inspection one can notice that the QFI is a function of $H_\theta=\hp_\theta+\hc$ and its first derivative. 
By defining $H':=\dot{H}_\theta$ we then solve the maximisation of the QFI~\eqref{eq:QFI_J} \emph{for given $H'$ over all possible $\hc$}, which is equivalent to maximise over all possible $H_\theta\equiv \hp_\theta+\hc$ (in the case of low temperature/gapped systems the same maximisation is performed over all possible $H_\theta$ with a non-degenerate ground state and a first excited gap $\Delta\gg \beta^{-1}$).

The results of such maximisation are therefore uniquely functions of $H'$. In particular our main results Eq.s~\eqref{eq:maxFI} and \eqref{eq:maxQFI_gap} only depend on the eigenvalues of $H'$, and are in particular proportional to its spectral norm $\|H'\|$ squared
\begin{align}
    \|H'\|=\lambda_{\rm M}-\lambda_{\rm m}\;.
\end{align}
It is thus important to notice that such bounds are \emph{always valid} once the Hamiltonian perturbation $\dot{H}_\theta$ and the temperature $(k_{\rm B}\beta)^{-1}$ (or, in the low temperature limit, the value of the gap $\Delta$) are given.

Secondly, by simple inspection of~\eqref{eq:maxFI} and \eqref{eq:maxQFI_gap} it is clear that if one is able to externally manipulate the spectral gap of $H$ such bound is straightforwardly modified.
In fact, in the most general decomposition of an externally-controlled $H_\theta$, both terms in~\eqref{eq:H_P+H_C} might depend on external tunable controls $\vec{y}$ as
\begin{align}
    H_\theta= \hp_\theta (\vec{y}) + \hc(\vec{y}) \;. 
\end{align}
However, without any restriction given to the degree of control on $\hp_\theta(\vec{y})$ one could in principle obtain infinite precision in the estimation of $\theta$, a situation which is unphysical. 
A trivial example can be provided by 
\begin{align}
H_\theta=y_1 \theta  H_1+ y_2 H_2\;.
\end{align}
In such case one should clearly use the highest allowed value of $|y_1|$ in order to scale the estimation error by $1/|y_1|$. Such $\max |y_1|$ needs to be bounded in any physical scenario. This is in fact a simple rescaling of the parameter to be measured, and is a feature shared by any metrological scheme.

In the case of a control-dependent $\hp_\theta$, the bounds provided in our work thus remain valid, (and can be in principle saturated when allowing all possible controls $\hc$), by simply choosing the $\hp_\theta(\vec{y})$ that in the specific control scenario features the maximum spectral norm $\max_{\vec{y}}\|\dot{H}^{\rm P}_\theta\|$.

For these reasons, in our main we restricted our discussion to the physically motivated scenario of~\eqref{eq:H_P+H_C}, that is formally
\begin{align}
    H_\theta=\hp_\theta+\hc(\vec{y})\;.
\end{align}


\subsection{Classical case}
In the classical case, all operators are diagonal in the same basis and therefore commute. In particular, all the superoperators introduce above coincide as $\J_{\rm B,\rho}[A] =\J_{\rm L,\rho}[A] =\mathcal{J}_{\rho}[A] =\rho A$ in this case.
It immediately follows that the Fisher Information~\eqref{eqapp:b-2_Fish} reduces to
\begin{align}
\label{eq:F_therm_clas}
   \beta^{-2} \F_\theta^{\rm diag}={\rm Var}_{\vec{p}_\theta}[\dot{\vec{H}}_\theta]
    =\sum_i p_i H'^2_{i} - \left(\sum_i p_i H'_{i}\right)^2
    =\beta^{-2}\sum_i \frac{\dot{p}^2_i}{p_i}\;.
\end{align}
$p_i$ being the populations of the state $\rho_\theta\equiv{\rm diag}(p_1,\dots,p_D)$, that is
\begin{align}
    p_i=\frac{e^{-\beta H_i}}{\sum_j e^{-\beta H_j}}
\end{align}
where we left implicit the dependence on the parameter $\theta$ in the energy levels.

\subsection{Explicit classical-quantum split}
\label{subsec:CQ_split}
Using the coordinate expressions~\eqref{eqapp:J_B_coordi} and~\eqref{eqapp:J_L_coordi}, one can write the action of $\mathcal{J}_\rho$ in components as
\begin{align}
    \mathcal{J}_\rho[A]=\sum_{ij} \frac{2(p_i-p_j)^2}{(\ln p_i-\ln p_j)^2(p_i+p_j)} A_{ij} \ketbra{i}{j}\;.
\end{align}
This expression apparently presents singularities for all terms satisfying $p_i=p_j$. It is however a matter of taking the limit $p_i\rightarrow p_j$ to see that such factor actually converges to $\frac{2(p_i-p_j)^2}{(\ln p_i-\ln p_j)^2(p_i+p_j)}\rightarrow p_i$.

More formally, consider the projectors $\Pi_a$ on the subspace of eigenvectors $\ket{i}$ of $\rho$ sharing the same eigenvalues (probability) $a$, i.e.
\begin{align}
    \Pi_a=\sum_{i|p_i=a} \ketbra{i}{i}\;.
\end{align}
The state can now be expressed as $\rho = \sum_a a \, \Pi_a$. It is straightforward to verify that
\begin{align}
    \J_{\rm B,\rho}[\Pi_a A \Pi_a]=\J_{\rm L,\rho}[\Pi_a A \Pi_a]=\mathcal{J}_{\rho}[\Pi_a A \Pi_a]=a \Pi_a A \Pi_a\;.
\label{eqapp:JPPaa}
\end{align}
while
\begin{align}
    \mathcal{J}_{\rho}[\Pi_a A \Pi_b]= \frac{2(a-b)^2}{(\ln a-\ln b)^2(a+b)}\Pi_a A \Pi_b\;.
\label{eqapp:JPPab}
\end{align}
Using properties~\eqref{eqapp:JPPaa} and~\eqref{eqapp:JPPab} we can then write
\begin{align}
    \Tr{A\mathcal{J}_\rho[A]}=\sum_{a} a \Tr{A\Pi_a A\Pi_a} 
    +\sum_{a\neq b} \frac{2(a-b)^2}{(\ln a-\ln b)^2(a+b)}\Tr{ A \Pi_b A \Pi_a}
\end{align}
This expression allows us to express the QFI~\eqref{eq:QFI_J}
as
\begin{align}
\label{eqapp:profish}
   \beta^{-2}\F_\theta= 
   \Tr{\sum_a a(\Pi_a H' \Pi_a)^2}-\Tr{\sum_a a\Pi_a H' \Pi_a}^2  +\sum_{a\neq b} \frac{2(a-b)^2}{(\ln a-\ln b)^2(a+b)}\Tr{ H' \Pi_b H' \Pi_a}\;.
\end{align}
We notice that the first term coincides with the classical variance of the block-diagonal perturbation $\sum_a \Pi_a H' \Pi_a$ on $\rho$. In each block $a$ we can choose the eigenvectors $\ket{i}$ of $\rho$ in order to diagonalize $\Pi_a H' \Pi_a$. By choosing such basis in all sectors $a$, we can rewrite Eq.~\eqref{eqapp:profish} as
\begin{align}
\label{eqapp:profish2}
   \beta^{-2}\F_\theta= 
   \sum_i p_i H'^2_{ii} - \left( \sum_i p_i H'_{ii} \right)^2 + \sum_{i,j| p_i\neq p_j} \frac{2(p_i -p_j)^2}{(\ln p_i - \ln p_j)^2(p_i+p_j)} |H'_{ij}|^2\;.
\end{align}
The first term represents the classical variance of the block-diagonal $H'$, which we dub $\F^{\rm diag}_\theta$ and represents the classical Fisher information obtained when measuring in the basis of $\rho_\theta$ (equivalently, $H_\theta$).
Finally, it is sufficient to rearrange the summation $p_i\neq p_j \rightarrow p_i > p_j$ (which is compensated by a factor 2), we arrive at the expression~\eqref{eq:QFI_components} of the main text, that is
\begin{align}
\label{eqapp:profish3}
    \F_\theta=\F_\theta^{\rm diag} + \sum_{i,j|p_i > p_j} \beta^2\frac{4(p_i-p_j)^2|H'_{ij}|^2}{(\ln p_i - \ln p_j)^2(p_i+p_j)} \;.
\end{align}
For full reconciliation to~\eqref{eq:QFI_components}, simply notice that for thermal states it holds $\ln p_i-\ln p_j=\beta(E_j-E_i)$.

\section{Inequalities}
\label{sec:ineq_derivations}
Here we show explicitly the two main inequalities used in the main text to obtain our main bound~\eqref{eq:maxFI}.

\subsection{Proof and saturation of inequality~\eqref{eq:main_ineq}}
The first inequality~\eqref{eq:main_ineq} follows from $\mathcal{J}_\rho \leq \J_{\rm B,\rho}$ as superoperators, and the following trivial identity $\Tr{H'\J_{\rm B,\rho}[H']}=\Tr{\rho H'^2}$. That is, \eqref{eq:main_ineq} can be written more explicitly as
\begin{align}
    \Tr{H'\mathcal{J}_{\rho_0}[H']} - \Tr{\rho_0 H'}^2 
\leq \Tr{H'\J_{\rm B,\rho_0}[H']} - \Tr{\rho_0 H'}^2 
= \text{Var}_{\rho_0}[H']\;.
\end{align}
The fact that $\mathcal{J}_\rho \leq \J_{\rm B,\rho}$ can be seen e.g. in coordinates, by
\begin{align}
\label{eq:ineq_coor}
     |A_{ij}|^2 \frac{2( p_i-\rho_j)^2}{(\ln p_i-\ln p_j)^2( p_i+p_j)} &\leq
    |A_{ij}|^2 \frac{p_i+p_j}{2}\;,
\end{align}
which holds term by term, due to the fact that for positive $x\geq 0$
\begin{align}
    \frac{2(1-x)^2}{(1+x)(\ln{x})^2}\leq \frac{1+x}{2}\;.
\end{align}
Therefore summing on $\{i,j\}$~\eqref{eq:ineq_coor} we obtain
\begin{align}
  \Tr{A\mathcal{J}_\rho[A]} &\leq \Tr{A\J_{\rm B,\rho}[A]} \quad\forall A\;,
\end{align}
which proves~\eqref{eq:main_ineq}.

\subsection{Upperbound and saturation of the maximum variance of any operator}
\label{app:std_var_ineq}
Here we give, for completeness, the maximum variance that can be attained when measuring a (bounded) observable on an arbitrary state.

Consider a bounded Hermitian operator $A$ with extremal eigenvalues $\lambda_{\rm M} \geq \lambda_{\rm m}$. Define another hermitian operator $\tilde A  = A -\mu \openone$ with $\mu =\frac{\lambda_{\rm M}+\lambda_{\rm m}}{2}$, whose extremal eigenvalues are $\pm \frac{\lambda_{\rm M}-\lambda_{\rm m}}{2}$. Since the spectra of both operators are related by a constant shift, their variance is the same and we have

\begin{equation}
    \text{Var}_\rho [A] =  \text{Var}_\rho [\tilde A] = \Tr{\rho \tilde A^2} -\Tr{ \rho \tilde A}^2 \leq \Tr{\rho \tilde A^2} \leq \| \tilde A^2\| = \left(\frac{\lambda_{\rm M}-\lambda_{\rm m}}{2} \right)^2.
\end{equation}
Finally, note that the bound is saturated by preparing $\rho =\frac{1}{2}\left( \ketbra{\lambda_{\rm M}}+\ketbra{\lambda_{\rm m}}\right)$, or any state with $\bra{\lambda_{\rm M}}\rho\ket{\lambda_{\rm M}}=\bra{\lambda_{\rm m}}\rho\ket{\lambda_{\rm m}}=1/2$.

\subsection{Upperbound and saturation of the maximum variance of any operator with a constrained state}
\label{app: variance}

Next we upper bound the variance in the case where the state is of the form $\rho = p_0 \ketbra{0} + (1-p_0)\rho_\perp$,
where $A \ket{0}= a_0 \ket{0}$ is the eigenstate of the operator and $p_0\geq 1/2$. Since the variance of an operator is unchanged by shifting its spectrum, we can take $A$ with extremal eigenvalues $\pm \frac{\lambda_{\rm M}-\lambda_{\rm m}}{2}\pm \delta$, and assume that $a_0\geq 0$ without loss of generality. 
To get the bound note that $\text{Var}_\rho[A] = \min_x \Tr{ \rho(A-x \openone )^2}$, which can be seen by deriving this expression wrt to $x$ and noticing that there is a single local (and global) minimum where $x=\Tr{\rho A }$. 
In our case we have
\begin{align}
 \text{Var}_\rho[A] &= \min_x \Tr{\rho(A-x \openone )^2 } \leq \min_{x\geq 0}  \Tr{\rho(A-x \openone )^2 }\\
 &\leq \min_{x\geq 0} \Big(p_0 (a_0-x)^2 + (1-p_0) \Tr{\rho_\perp (A-x\openone )^2} \Big) 
 \\
 &\leq \min_{x\geq 0} \left(p_0 (a_0-x)^2 + (1-p_0) (x+\delta)^2 \right),
\end{align}
since in the last inequality $\Tr{\rho_\perp (A-x\openone )^2}\leq \|(A-x\openone)^2\| = (\delta +x)^2$ for $x\geq 0$. We now set $x$ to the following positive value
$ x = \begin{cases}
    p_0 a_0 -(1-p_0) \delta & a_0 \geq \frac{1-p_0}{p_0} \delta\\
    0 & a_0 < \frac{1-p_0}{p_0} \delta
    \end{cases}$ to obtain the bound
\begin{align}
    \text{Var}_\rho[A] &\leq \begin{cases}
    p_0(1-p_0)(a_0+\delta)^2 & a_0 \geq \frac{1-p_0}{p_0} \delta\\
    p_0 a_0^2 + (1-p_0) \delta^2 & a_0 < \frac{1-p_0}{p_0} \delta
    \end{cases}\quad
    \leq 
    \quad
    \begin{cases}
    p_0(1-p_0) (2\delta)^2 & a_0 \geq \frac{1-p_0}{p_0} \delta\\
    \frac{1-p_0^2}{p_0} \delta^2 + (1-p_0) \delta^2 & a_0 < \frac{1-p_0}{p_0} \delta
    \end{cases}  \leq  p_0(1-p_0) (\lambda_{\rm M}-\lambda_{\rm m})^2.
\end{align}
Where for the last inequality we used $ \frac{1-p_0^2}{p_0} \delta^2 + (1-p_0) \delta^2  = \frac{p_0(1-p_0)}{p_0^2}\delta^2 \leq p_0(1-p_0) 4 \delta^2 = p_0(1-p_0) (\lambda_{\rm M}-\lambda_{\rm m})^2$ guaranteed by $p_0\geq 1/2$.  Finally, note that the bound 
\begin{equation}
      \text{Var}_\rho[A]  \leq  p_0(1-p_0) (\lambda_{\rm M}-\lambda_{\rm m})^2
\end{equation}
is saturated by setting $\ket{0}= \ket{\lambda_{\rm M}}$ and preparing  $\rho = p_0 \ketbra{\lambda_{\rm M}} + (1-p_0) \ketbra{\lambda_{\rm m}}$.

Consider now a thermal state defined by
\begin{align}
    \rho=\frac{e^{-\beta \hc}}{\Tr{e^{-\beta\hc}}}
\end{align}
including a gap $\Delta$ between the ground state and the the first excited level. In the low temperature limit the probability to find the system outside of the ground state decays at least as 
$1-p_0 \leq (D-1)e^{-\beta \Delta}$ where $D$ is the dimension of the Hilbert space, leading to the worst case bound
\begin{align}
    \text{Var}_\rho[A]\leq \beta^2 (D-1)e^{-\beta \Delta}(\lambda_{\rm M}-\lambda_{\rm m})^2.
\end{align}
Saturating this inequality requires \textit{both} $A$ and $\hc$ to have a $(D-1)$ degenerate eigenstate: of energy $\lambda_{\rm m}$ or $\lambda_{\rm M}$ for $A$, and $E_i=\Delta$ for all states above ground energy in $\hc$. This is quite exotic, and in general one expects a much faster decay of $\text{Var}_\rho[A]$ in the low temperature limit. These considerations apply in bounding the classical contribution $\F_\theta^{\rm diag}$~\eqref{eq:QFI_components} to the Fisher information (by substituting $A=H'$).
\\

Finally, let us briefly comment on the fact that the eigenvalues of a dephased operator $\tilde A = \sum_a \Pi_a A \Pi_a$ are contained between the extremal eigenvalues $\lambda_{\rm M}\geq \lambda_{\rm m}$ of $A$. To see this, simply notice that for any state $\ket{\psi}$ 
\begin{equation}
    \bra{\psi} \tilde A \ket{\psi} = \sum p_a \bra{\psi_a} A \ket{\psi_a}
\end{equation}
where $p_a = \| \Pi_a \ket{\psi} \|^2 $ and $\ket{\psi_a} = \frac{\Pi_a \ket{\psi}}{\sqrt{p_a}}$. Since $\lambda_{\rm M} \geq \bra{\psi_a} A \ket{\psi_a} \geq \lambda_{\rm m}$ we conclude that
\begin{equation}
    \lambda_{\rm M} \geq \bra{\psi} \tilde A \ket{\psi} \geq \lambda_{\rm m}.
\end{equation}

\section{ {A note on dynamical metrology and analogies with new results}}
\label{app:dyna_compa}

We refer in this section to the standard framework of \emph{dynamical metrology} in its simplest scenario, that is we consider a unitary parameter-encoding given by free evolution under $H_\theta$,  of the form
\begin{align}
\label{eq:dynamical_encod}
    \rho_\theta=U_\theta\psi U_\theta^\dagger \equiv e^{-i H_\theta t/\hbar}\psi e^{i H_\theta t/\hbar}\;,
\end{align}
where $\psi:=\ketbra{\psi}{\psi}$ is an initially prepared state that we consider here to be pure. In fact, as the QFI is a convex quantity, when being interested in its maximum value over all possible initial preparations $\psi$, the latter can be taken to be pure without loss of generality.
The QFI of $\rho_\theta$~\eqref{eq:dynamical_encod} w.r.t. $\theta$ can then be computed via standard techniques~\cite{pang2014quantum} to be
\begin{align}
    \F_\theta=4 {\rm Var}_\psi [i(\partial U_\theta) U_\theta^\dagger]\;.
\end{align}
Notice now that the argument of such variance can be directly computed as~\cite{pang2014quantum,hiai2014introduction}
\begin{align}
\label{eq:idUUdag}
    i(\partial U_\theta) U_\theta^\dagger=\int_0^1 \D s\; e^{-i s H_\theta t/\hbar}\frac{t}{\hbar}\partial H_\theta e^{i s H_\theta t/\hbar} := \int_0^1 \D s\;\frac{t}{\hbar} H'^{(s)}_\theta \;.
\end{align}
This expression can thus be seen as an average of partially-rotated Hamiltonian derivatives $H'^{(s)}_\theta$.
It then follows immediately by convexity that
\begin{align}
    \max_\psi{\rm Var}_\psi [i(\partial U_\theta) U_\theta^\dagger]\leq \frac{t^2}{\hbar^2} \max_s \max_\psi {\rm Var}_\psi [H'^{(s)}_\theta]\;.
\end{align}
However by unitary invariance 
\begin{align}
    \max_\psi {\rm Var}_\psi [H'^{(s)}]\equiv \max_\psi {\rm Var}_\psi [\partial_\theta H_\theta]\quad \forall s\;.
\end{align}
For which we can state the variance-based bound
\begin{align}
\label{eq:dynam_bound_var}
    \F_\theta \leq 4 \frac{t^2}{\hbar^2} \max_\psi {\rm{Var}}[\partial_\theta H_\theta]\;.
\end{align}
We notice as well that such bound can be saturated whenever $[H_\theta,\partial_\theta H_\theta]=0$, which implies in~\eqref{eq:idUUdag} $i(\partial U_\theta) U_\theta^\dagger\equiv \frac{t}{\hbar}\partial_\theta H_\theta$ in such case.
In particular, assuming a control of the form $H_\theta=\hp_\theta+\hc$ as from~\eqref{eq:H_P+H_C} (cf. also \ref{sec:local_control_rationale}) such bound can always be saturated.

Equation~\eqref{eq:dynam_bound_var} thus provides an expression which is useful to drive analogy with the main results of our work, which characterise the maximum value of the QFI $\F_\theta$ for two other canonical encodings of the Hamiltonian parameter $\theta$ -- namely, in the thermal state~\eqref{eq:thermal_state} or in the ground state of $H_\theta$ itself --  in terms of a variance maximisation, see Table~\ref{tab:variance_compa}.

\begin{table}[]
    \centering
    \begin{tabular}{c|c}
 \sf{Encoding} & \sf{Maximum} $\F_\theta$ 
 \\ \hline \rule{0pt}{4ex} 
  \text{Dynamical}  $\rho_\theta = e^{-i H_\theta t/\hbar}\psi  e^{i H_\theta t/\hbar}$ & $
  \dfrac{4t^2}{\hbar^2} \max_\psi {\rm{Var}}_{\psi}[\partial_\theta H_\theta]
  $ \rule[-2.5ex]{0pt}{0pt}  \\ \hline \rule{0pt}{4ex}    
 \text{Thermal} $ \rho_\theta= \dfrac{e^{-\beta H_\theta}}{\Tr{e^{-\beta H_\theta}}} $ & $\beta^2 \max_{\rho_\theta} {\rm{Var}}_{\rho_\theta}[\partial_\theta H_\theta]$ \rule[-2.5ex]{0pt}{0pt}\\ \hline \rule{0pt}{4ex} 
 Ground $\ket{g_\theta}$ of $H_\theta$ with minimum gap $\Delta$ to first excited & $\dfrac{4}{\Delta^2} \max_{\ket{g}} {\rm{Var}}_{\ket{g}}[\partial_\theta H_\theta]$ \rule[-2.5ex]{0pt}{0pt} \\ \hline
\end{tabular}
    \caption{ {Canonical encodings of Hamiltonian parameters and their corresponding maximum Quantum Fisher Information $\F_\theta$, in terms of variance maximisation on the unperturbed state. The same table is presented in the main text, by substituting the identity $\max {\rm Var}[A]=\|A\|^2/4$, which is valid for any operator $A$, $\|A\|:=\lambda_{\rm M}-\lambda_{\rm m}$ being the seminorm defined by the difference of maximum and minimum eigenvalues of $A$.}}
    \label{tab:variance_compa}
\end{table}

\section{SLD and optimal measurement}
\label{app:SLD_and_meas} 

As well known in quantum metrology the QFI can be saturated by measuring the state $\rho_\theta$ in the eiegenbasis of the symmetric logarithmic derivative operator (SLD)~\cite{paris2009quantum}, given by 
\begin{align}
    \J_{\rm B,\rho_\theta}^{-1}[\dot{\rho}_\theta]\bigg|_{\theta=0}
    =\sum_{ij}\frac{2 (p_i-p_j)}{(\ln p_i-\ln p_j)(p_i+p_j)} H'_{ij} \ketbra{i}{j}
    =\sum_{ij}\frac{2 \tanh{(\beta(E_i-E_j)/2)}}{\beta (E_i-E_j)}H'_{ij} \ketbra{i}{j}\;.
\end{align}
In this expression we used $\rho_0\propto e^{-\beta \hc}$ and thus $p_{i}\propto e^{-\beta E_i}$. Moreover, as detailed in Sec.~\ref{app:analytical}, we slightly abused the notation as for $p_i=p_j$ ($E_i=E_j$) the terms in the sum are seemingly singular. The singularity is removed by considering the appropriate limit $p_i\rightarrow p_j$ ($E_i\rightarrow E_j$), or following the same steps presented in Sec.~\ref{subsec:CQ_split}
and formally
\begin{align}
    \J_{\rm B,\rho_\theta}^{-1}[\dot{\rho}_\theta]\bigg|_{\theta=0}= \sum_{E} \Pi_E H'\Pi_E + \sum_{ij|E_i\neq E_j}\frac{2 \tanh{(\beta(E_i-E_j)/2)}}{\beta (E_i-E_j)}H'_{ij} \ketbra{i}{j}\;,
\end{align}
$\Pi_E$ being the projectors on the sub-spaces of energy $E$ of $\hc$.

Clearly if the relevant Hamiltonians commute $[\hc,H']=0$ the optimal measurement basis coincides with any choice that diagonalizes both, $\hc, H',\rho_\theta$, in compatibility with the fact that in the classical case the choice of measurement basis is trivial. In general, the diagonal basis of the SLD does not coincide with that of $\hc$, nor $H'$.
In the following we comment on relevant cases.

\paragraph{Measuring in the basis of $\hc$.}
When choosing to measure the system in the basis of $\hc$, the corresponding Fisher Information is given by $\F^{\rm diag}_\theta$  in Eqs~\eqref{eqapp:profish2}-\eqref{eqapp:profish3}, which is clearly smaller than the total QFI as expected.

\paragraph{Measuring $H'$.}
The expression of the Fisher Information relative to the measurement basis that diagonalizes $H'$ is non trivial to obtain in general.
However, for a comparison, consider the estimation of $\theta$ based on the average value of the perturbation $H'$. In the limit of large $n$ repetitions, the error will be approximated by
\begin{align}
    n^{-1}\langle\Delta{\theta}^2\rangle^{-1} &\approx {\rm Var}_{\rho_0}[H']^{-1}\left(\frac{\partial\langle H'\rangle}{\partial\theta}\right)^2
\end{align}
which corresponds to an effective Fisher Information $\tilde{F}_\theta$ which is given by, choosing by gauge freedom $\Tr{H'\rho_0}=0$,
\begin{align}
\label{eqapp:Ftilde}
    \tilde{\F}^{\langle H' \rangle}_\theta &\approx \beta^2 \frac{\Tr{H'\J_{\rm L,\rho_0}[H']}^2}{\Tr{H'^2\rho_0}}\;.
\end{align}
By Cauchy-Schwarz inequality we see that this expression is in general smaller than $\beta^2\Tr{H'\mathcal{J}_{\rho_0}[H']}$ corresponding to the total QFI, as expected. 

\paragraph{Low temperature.}
In the low temperature limit, the maximal attainable QFI ~\eqref{eq:maxQFI_gap} described in the main text, can be obtained by measuring the system in the base of the perturbation $H'$. To see this, consider the optimal configuration described in the main text to saturate~\eqref{eq:maxQFI_gap}, that is
\begin{align}
    \hc=\epsilon\ketbra{\psi^+}{\psi^+}+(\epsilon+\Delta)\ketbra{\psi^-}{\psi^-}+H_\perp\;,
\end{align}
with $H_\perp > \epsilon+\Delta$ and
\begin{align}
       \ket{\psi^{\pm}}:=\frac{\ket{\lambda_{\rm M}}\pm\ket{\lambda_{\rm m}}}{\sqrt{2}}\;.
\end{align}
Then it is easy to verify that in the limit $\beta\rightarrow\infty$ the expression $\tilde{\F}^{\langle H' \rangle}_\theta$ saturates~\eqref{eq:maxQFI_gap}. Indeed without loss of generality we offset $H'$ such that $\lambda_{\rm M}=-\lambda_{\rm m}:=\lambda>0$, ensuring $\Tr{H'\rho_0}=0$, and compute~\eqref{eqapp:Ftilde}
\begin{align}
    \beta^2\frac{\Tr{H'\J_{\rm L,\rho_0}[H']}^2}{\Tr{H'^2\rho_0}} \approx \beta^2 \left(2\frac{|H'_{+-}|^2}{\beta \Delta}\right)^2\frac{1}{\lambda^2}=\frac{4\lambda^4}{\Delta^2\lambda^2}=\frac{(\lambda_{\rm M}-\lambda_{\rm m})^2}{\Delta^2}\;,
\end{align}
which coincides with~\eqref{eq:maxQFI_gap}.
In the computation, we made use of the fact that the term
\begin{align}
    \frac{p_i-p_j}{\ln p_i-\ln p_j} |H'_{ij}|^2
\end{align}
is nonzero only for $\{i,j\}=\{+,-\}$, as $H'_{--}=H'_{++}=0$ and $\frac{p_i-p_j}{\ln p_i-\ln p_j}$ is exponentially suppressed for higher energy levels.

\section{Proof of $\J_{{\rm B},\eye+\varepsilon}-\mathcal{J}_{\eye+\varepsilon}\approx \mathcal{O}(\varepsilon^2)$}
\label{app:J_loc_eye}

The $\J_{\rm B,\rho}$, $\J_{\rm L,\rho}$ and $\mathcal{J}_{\rho}$ superoperators all coincide, at first order, when $\rho$ is close to a flat distribution. That is
\begin{align}
    \rho=\alpha(\eye+\epsilon) \Rightarrow \mathcal{J}_{\rho}\approx \J_{\rm L,\rho} \approx \alpha\J_{\rm B,\eye+\epsilon}+\mathcal{O}(\epsilon^2)\;,
\end{align}
where $\alpha>0$ is any positive scalar.
In order to verify this property, it is sufficient to explicitly evaluate for $x:=p_j/p_i\approx 1+\epsilon$\;,
\begin{align}
    \mathcal{J}_\rho[A]_{ij}= \frac{2(p_i-p_j)^2}{(\ln p_i -\ln p_j)^2(p_i+p_j)} A_{ij}=p_i A_{ij}\frac{2(1-x)^2}{(\ln x)^2(1+x)}\approx p_i\frac{(1+x)}{2} A_{ij}=\frac{p_i+p_j}{2} A_{ij}\;,
\end{align}
where we used that the function $\frac{2(1-x)^2}{(\ln 1-\ln x)^2(1+x)}$ can be approximated at first order by $\frac{1+x}{2}$ when expanded around $x=1$.
The same approximation holds for
\begin{align}
    \J_{\rm L,\rho}[A]_{ij}=\sum \frac{p_i-p_j}{\ln p_i-\ln p_j} A_{ij}\approx \frac{p_i+p_j}{2} 
\end{align}
in the same limit.
More in general, the property here presented is valid for all $\J$ superoperators belonging to the Petz family~\cite{petzMonotoneMetricsMatrix1996} of generalized Quantum Fisher Informations~\cite{scandi2023quantum}, as they all share the property of being parametrised by functions $f:\mathbf{R}^+\rightarrow\mathbf{R}^+$ as
\begin{align}
    \J_{f,\rho}:=f(\LL_\rho\circ\RR^{-1}_\rho)\circ \RR_\rho \quad \text{where} \quad \RR_\rho[A]:=A\rho\;,\; \LL_\rho[A]:=\rho A\;,
\end{align}
which in components reads
\begin{align}
    \J_{f,\rho}[A]_{ij}=f(p_i/p_j)p_j A_{ij}\;.
\end{align}
The property of all $\J_{f}$ locally coinciding in the is recovered by the fact that they all satisfy
\begin{align}
    f(1)=1\quad\text{and}\quad f'(1)=\frac{1}{2}\;.
\end{align}
and therefore $f(x)\approx \frac{1+x}{2}$ around $x\approx 1$\;.

%

 \section{1-D Ising chain example}
\label{app: Ising}
We now consider specifically the classical 1D Ising model described by the Hamiltonian
\begin{align}\label{app: ID ising H}
    H_\theta = - J \sum_{i=1}^N Z_i Z_{i+1} + (B+\theta_Z) \sum_{i=1}^N Z_i + \theta_X \sum_{i=1}^N X_i 
\end{align}
with nearest neighbour interactions, periodic boundary conditions $Z_{N+1}= Z_1$. We will compute the QFI of the resulting Gibbs state for the parameters $\theta_Z$ and $\theta_X$ at $\theta_Z=\theta_X=0$, where the Hamiltonian is easy to diagonalize. Then we discuss the behaviour of the resulting expressions in the "thermodynamic" ($N\to \infty$) and low temperature ($\beta\to \infty$) limits.

\subsection{Computing $\F_{\theta_Z}$}

We start by computing the QFI with respect to the parallel field $\F_{\theta_Z}$. At $\theta_X=0$ the model $H_{\theta_Z}$ is classical, i.e. diagonal in the same (computational) basis for all values of $B,J$ and $\theta_Z$. For a classical model ($[H,H']=0$) in the linear regime $H_\theta = H + \theta H'$, it is well known that the QFI of the Gibbs state is related to the partition function $\mathcal{Z}_\theta = \Tr{ e^{-\beta H_\theta}}$ as
\begin{equation}
    \mathcal{F}_\theta = \frac{\partial^2} {\partial \theta^2}  \ln \mathcal{Z}_\theta.
\end{equation}
For completeness, let us quickly show this relation. We have
\begin{align}
     \frac{\partial^2} {\partial \theta^2}  \ln \mathcal{Z}_\theta =  \frac{\mathcal{Z}''_{\theta}}{\mathcal{Z}_\theta} -  \left( \frac{\mathcal{Z}'_{\theta}}{\mathcal{Z}_\theta} \right)^2 = \frac{\Tr{  -\beta \frac{\partial}{\partial \theta}e^{-\beta H_\theta } H'}}{\mathcal{Z}_\theta} 
     - 
\left(-\beta \frac{ e^{-\beta H_\theta } H'}{\mathcal{Z}_\theta}\right)^2
     = \beta^2\left (\Tr{ \rho_\theta (H')^2} - \Tr{ \rho_\theta H'}^2 \right) = \mathcal{F}_\theta,
\end{align}
where the last equality is discussed in the main text.

It is also well known that the partition function of the classical 1D Ising chain can be computed analytically. As a warm-up we will now repeat this derivation, since the computation of $\F_{\theta_X}$ will be very similar. The basis idea is to express the partition function as a trace of a large product of matrices. To shorten the notation we now absorb $\theta_Z$ into $B$ and simply write
\begin{align}
    \mathcal{Z} = \Tr{e^{-\beta H}} =   \sum_{z_1,\dots,z_N=\pm 1} e^{ \beta J z_{N} z_1  } e^{-\beta  B z_1}  e^{\beta J z_1 z_2} \dots e^{\beta J z_{N-1} z_N} e^{-\beta z_N}.
\end{align}
Introducing the matrices 
\begin{equation} \label{eq: matrices A C}
A= \left(\begin{array}{cc} 
e^{-\frac{1}{2}\beta B}& \\
& e^{\frac{1}{2}\beta B}
\end{array}
\right)\qquad
\text{and} \qquad C= \left(\begin{array}{cc} 
e^{\beta J}& e^{-\beta J}\\
e^{-\beta J}& e^{\beta J}
\end{array}
\right)
\end{equation}
allows one to rewrite the partition function as
\begin{align}
    \mathcal{Z} &=\sum_{z_1,\dots,z_N=1,2} (A C A)_{z_N z_1} (A C A)_{z_1 z_2} \dots (A C A)_{z_{N-1} z_N} 
     = \Tr {(ABA)^N} = \lambda_+^N + \lambda_-^N.
\end{align}
Where $\lambda_{\pm}$ are the eigenvalues of the matrix $ACA$ given by
\begin{equation} \label{eq: lambda pm} 
\lambda_{\pm} = e^{\beta J} \left(\cosh(\beta B) \pm  \sqrt{\sinh^2(\beta B) + e^{- 4 \beta J}} \right). 
\end{equation}
The QFI for a parallel field thus reads
\begin{equation} \label{app: QFIZ der}
 \mathcal{F}_{\theta_Z} = \frac{\partial^2}{\partial B^2} \ln \left( \lambda_+^N + \lambda_-^N \right).
\end{equation}

\subsection{Computing $\F_{\theta_X}$. Step 1.}

Next, let us compute the QFI expression for transverse field. We set $\theta_Z=0$ and $\theta=\theta_X$ for short, the Hamiltonian reads 
\begin{equation}
H_\theta = - J \sum_{i=1}^N Z_i Z_{i+1} + B \sum_{i=1}^N Z_i + \theta \sum_{i=1}^N X_i.
\end{equation}
At $\theta=0$ we have $\rho = \frac{e^{-\beta H_0}}{\mathcal{Z}}$ and $H' = \sum_{i=1}^N X_i$. By linearity $ \J_{\rm L,\rho}[H'] = \sum_{i=1}^N  \J_{\rm L,\rho}[X_i]  $ and the individual terms in the sum are given by
\begin{align}
    \J_{\rm L,\rho}[X_i] = \int_{0}^1 \dd s \, \rho^{s} X_i  \rho^{1-s} = \frac{1}{\mathcal{Z}}\int_{0}^1 \dd s \, e^{-\beta s H_0} X_i  e^{-\beta (1-s) H_0}.
\end{align}
Let us decompose $H_0 = Z_i R_i + H_{\lnot i}$ with $R_i= (B - J (Z_{i-1}+Z_{i+1}))$ and $H_{\lnot i} = H_0 -R_i$. Note that $[R_i,H_{\lnot i}]= [X_i,H_{\lnot i}]=0$, allowing us to write
\begin{align}
    \J_{\rm L,\rho}[X_i] & = \frac{e^{-\beta H_{\lnot i}}}{\mathcal{Z}}\int_{0}^1 \dd s \, e^{-\beta s Z_i R_i} X_i  e^{-\beta (1-s) Z_i R_i}\\ 
    & = \frac{e^{-\beta H_{\lnot i}}}{\mathcal{Z}}\left( \int_{0}^1 \dd s \, e^{-\beta s Z_i R_i} \ketbra{0}{1}_i  e^{-\beta (1-s) Z_i R_i} + \text{h.c.} \right) \\
     & = \frac{e^{-\beta H_{\lnot i}}}{\mathcal{Z}}\left( \int_{0}^1 \dd s \, e^{-\beta s R_i} \ketbra{0}{1}_i  e^{\beta (1-s) R_i )} + \text{h.c.} \right)\\
     & = \frac{e^{-\beta H_{\lnot i}}}{\mathcal{Z}}\left( \ketbra{0}{1}_i \int_{0}^1 \dd s \, e^{\beta (1-2s) R_i}   + \text{h.c.} \right)\\
     & =  \frac{e^{-\beta H_{\lnot i}}}{\mathcal{Z}} \frac{\sinh(\beta R_i)}{\beta R_i} X_i
\end{align}
Next, using the analytical expression of $ \J_{\rm B,\rho}^{-1}$ given below Eq.\eqref{eq: SLD} we compute
\begin{align}
    \J_{\rm B,\rho}^{-1} \circ \J_{\rm L,\rho}[X_i]  &= 2 \int_0^\infty \dd s \; e^{-\rho s} \, \J_{\rm L,\rho}[X_i] \, e^{-\rho s} \\
    & =  \frac{e^{-\beta H_{\lnot i}}}{\mathcal{Z}} \frac{\sinh(\beta R_i)}{\beta R_i} \, 2\int_0^\infty \dd s \; e^{-\rho s} \, X_i\, e^{-\rho s} \\
    & =  \frac{e^{-\beta H_{\lnot i}}}{\mathcal{Z}} \frac{\sinh(\beta R_i)}{\beta R_i}\,  2\int_0^\infty \dd s \; \exp( -s \frac{e^{-\beta H_{\lnot i}}e^{-\beta Z_i R_i}}{\mathcal{Z}}) \, X_i\, \exp( -s \frac{e^{-\beta H_{\lnot i}}e^{-\beta Z_i R_i}}{\mathcal{Z}}) \\
    & =  \frac{e^{-\beta H_{\lnot i}}}{\mathcal{Z}} \frac{\sinh(\beta R_i)}{\beta R_i} X_i 2 \int_0^\infty \dd s \; \exp( -s \frac{e^{-\beta H_{\lnot i}}(e^{-\beta R_i}+e^{+\beta R_i})}{\mathcal{Z}}) \\
    &= \frac{e^{-\beta H_{\lnot i}}}{\mathcal{Z}} \frac{\sinh(\beta R_i)}{\beta R_i} X_i \frac{\mathcal{Z}}{e^{-\beta H_{\lnot  i}} \cosh(\beta R_i)}
    \\
    &=\frac{\tanh(\beta R_i)}{\beta R_i}  X_i
\end{align}
Finally, using Eq.~\eqref{eq:QFI_J} and $\Tr{\rho X_i}=0$ we obtain the QFI
\begin{align}
\F_{\theta_X} &= \beta^2     \Tr {\sum_{i,j} \J_{\rm L,\rho}[X_j] \J_{\rm B,\rho}^{-1} \circ \J_{\rm L,\rho}[X_i] }  \\
&= \frac{2}{\mathcal{Z}} \sum_i \Tr {e^{-\beta H_{\lnot i}} \frac{\sinh(\beta R_i)\tanh(\beta R_i)}{R_i^2}} \label{eq: QFIX exp}\\
&=  2\sum_i \Tr {\frac{e^{-\beta H_{0}}}{\mathcal{Z}} 2 \frac{\sinh(\beta R_i)\tanh(\beta R_i)}{\exp(-\beta Z_i R_i) R_i^2}}\\
& = N \, \mathds{E}\left[ 2 \frac{\sinh^2(\beta R_i)\exp(\beta Z_i R_i)}{\cosh(\beta R_i) R_i^2} \right],
\end{align}
where we used $\Tr{X_i X_i}=2$ and the fact that $\Tr{ X_i A X_j B}=0$ for any diagonal (in the computational basis) operators $A,B$ and  $i\neq j$.\\

To gain some physical intuition let us denote the function appearing in the expected value as 
\begin{equation}
\zeta(z_{i-1},z_i,z_{i+1}) = 2 \frac{\sinh^2\big(\beta (B- J (z_{i-1}+z_{i+1}))\big)\exp\big(\beta z_i (B- J (z_{i-1}+z_{i+1}))\big)}{\cosh\big(\beta (B- J (z_{i-1}+z_{i+1}))\big) (B- J (z_{i-1}+z_{i+1}))^2}. 
\end{equation}
For ferromagnetic interactions $J> 0$ with a positive field $B>0$ and in the limit of low temperature $\beta \to \infty$ one finds that
\begin{equation}
    \zeta(\downarrow\downarrow\downarrow) = \frac{1}{ (B+2J)^2} \implies \lim_{\beta\to\infty} \F_{\theta_X} = \frac{N}{ (B+2J)^2},
\end{equation}
which is consistent with the low-temperature limit discussed in the main text.

\subsection{Computing $\F_{\theta_X}$. Step 2.}

Let us now compute the expression in the Eq.~\eqref{eq: QFIX exp}. This can be done analogously to the partition function 
\begin{equation}\label{eq app: QFI X}
\F_{\theta_X}
= \frac{ 2}{\mathcal{Z}} \sum_{i=1}^N \Tr {e^{-\beta H_{\lnot i}} \frac{\sinh(\beta R_i)\tanh(\beta R_i)}{R_i^2}} = \frac{ 2 N }{\mathcal{Z}}  \Tr {e^{-\beta H_{\lnot 1}} \frac{\sinh^2(\beta R_1)}{\cosh(\beta R_1) R_1^2}},
\end{equation}
with $R_1 = B - J (Z_2+Z_N)$. To compute it, in addition to the matrices $A$ and $C$ in Eq.~\eqref{eq: matrices A C} introduce another matrix
\begin{equation}
\F_{\theta_X} = \left( 
\begin{array}{cc}
\frac{\sinh^2(\beta (B-2J))}{ (B-2J)^2 \cosh(\beta (B-2J)) } & \frac{\sinh^2(\beta B)}{ B^2 \cosh(\beta B)}\\
\frac{\sinh^2(\beta B)}{ B^2 \cosh(\beta B)}& \frac{\sinh^2(\beta (B+2J))}{ (B+2J)^2 \cosh(\beta (B+2J))}
\end{array}
\right)
\end{equation}
and express
\begin{align}
    \mathcal{F}_{\theta_X} &= \frac{2 N}{\mathcal{Z}} \sum_{z_2,\dots z_N=1,2} (A C A)_{z_2 z_3} \dots (A C A)_{z_{N-1}z_N} (A F A)_{z_{N}z_2} = \frac{2 N}{\mathcal{Z}} \Tr{ (ACA)^{N-2} AFA} \\
\end{align}
By diagoanlizing $(ACA)= \lambda_+ \bm v_+ \bm v_+^T + \lambda_- \bm v_- \bm v_-^T$ with $\lambda_{\pm}$ given in the Eq.~\eqref{eq: lambda pm} and
\begin{equation}
    \bm v_\pm = \frac{1}{\sqrt{ \mu_\pm^2+1}} \binom {\mu_\pm}{1} \qquad \text{with} \qquad \mu_\pm = - e^{2\beta J} \sinh(\beta B) \pm \sqrt{e^{4 \beta J} \sinh^2(\beta B)+1},
\end{equation}
we find the final expression for the QFI
\begin{equation}
  \mathcal{F}_{\theta_X} =2 N \frac{\lambda_+^{N-2}  \bm v_+^T (A F A) \bm v_+ + \lambda_-^{N-2}  \bm v_-^T (A F A) \bm v_-}{\lambda_+^N+\lambda_-^N}.
\end{equation}

\subsection{The limiting behavior of $\F_{\theta_Z}$ for $J>0$ and $B=0$.}

At zero magnetic field the expression of the QFI in Eq.~\eqref{app: QFIZ der} becomes particularly simple, direct calculation gives
\begin{equation}
B=0 :  \qquad \F_{\theta_Z} = \beta ^2 N e^{2 \beta  J} \frac{\cosh ^N(\beta  J)-\sinh ^N(\beta  J)}{\cosh ^N(\beta  J)+\sinh ^N(\beta  J)}.
\end{equation}
In the thermodynamic ($N\to \infty$) and zero temperature ($\beta\to 0$) limits this expression becomes
\begin{align}
B=0 \quad \text{and} \quad  N\to \infty: \qquad \F_{\theta_Z} & = N \beta^2 e^{2 \beta J} \\
B=0  \quad \text{and} \quad  \beta \, \to \infty: \qquad \F_{\theta_Z} & = N^2 \beta^2  (1- \frac{1}{3}(N^2-1)e^{-4 \beta J}) + \mathcal{O}(e^{-8 \beta J})
\end{align}
showing that the Heisenberg scaling is attainable, but requires that $\beta J$ increases logarithmically with $N$.\\

\subsection{The limiting behavior of $\F_{\theta_Z}$ for $J>0$ and $B>0$.}

With magnetic field the expression of the QFI in Eq.~\eqref{app: QFIZ der} is more cumbersome. Nevertheless we can write 
\begin{align}
    \F_{\theta_Z} &= \frac{\partial^2} {\partial B^2}  \ln \mathcal{Z}= \frac{\partial^2} {\partial B^2}  \ln\left[\left(\cosh(\beta B) + \sqrt{\sinh^2(\beta B) + e^{- 4 \beta J}} \right)^N +\left(\cosh(\beta B) -\sqrt{\sinh^2(\beta B) + e^{- 4 \beta J}} \right)^N \right] \\
    &=\frac{\partial^2} {\partial B^2}  \Big(\ln \cosh^N(\beta B) + \ln \left[\left(1+ \sqrt{1- \frac{1-e^{- 4 \beta J}}{\cosh^2(\beta B)}} \right)^N +\left(1 -\sqrt{1- \frac{1-e^{- 4 \beta J}}{\cosh^2(\beta B)}} \right)^N \right] \Big) \\
    & =\frac{\partial^2} {\partial B^2}  \Big(N \ln \cosh(\beta B) + \ln \left[\left(1+ G_B \right)^N +\left(1 - G_B\right)^N \right] \Big)
\end{align}
where we introduced the function $G_B = \sqrt{1- \frac{1-e^{- 4 \beta J}}{\cosh^2(\beta B)}}$ to shorten the notation. Here, for the first term one easily gets
\begin{equation}
   \F_{\theta_Z}^{(1)}= N \frac{\partial^2} {\partial B^2}  \ln \cosh(\beta B) = \frac{\beta^2 N}{\cosh^2(\beta B)}
\end{equation}
To compute the rest it is convenient to split the remaining terms in two parts as follows
\begin{align}
    \frac{\partial^2} {\partial B^2} \ln \left[\left(1+ G_B \right)^N +\left(1 - G_B \right)^N \right] = \frac{\partial} {\partial B} \frac{N \big(\left(1 + G_B \right)^{N-1} + \left(1 - G_B \right)^{N+1}\big) G'_B}{\left(1+ G_B \right)^N +\left(1 - G_B \right)^N} 
    = \F_Z^{(2)} + \F_Z^{(3)}
\end{align}
where applying the derivative againe one gets for the the two terms
\begin{align}
    \F_{\theta_Z}^{(2)} &=N^2 (G'_B)^2 \frac{\left(\left(1+ G_B \right)^{N-2} +\left(1 - G _B\right)^{N-2}\right)\left(\left(1+ G_B \right)^N +\left(1 - G_B \right)^N\right) -\left(\left(1+ G _B\right)^{N-1} +\left(1 - G_B \right)^{N-1}\right)^2}{\left(\left(1+ G_B \right)^N +\left(1 - G_B\right)^N\right)^2} \\
    & = N^2  \frac{4 (G'_B)^2 Gv^2 (1-G_B^2)^{N-2}}{\left(\left(1+ G _B\right)^N +\left(1 - G_B \right)^N\right)^2} = N^2  \frac{4 (G'_B)^2 G_B^2 (1-G_B^2)^{N-2}}{\left(\left(1+ G_B \right)^N +\left(1 - G _B \right)^N\right)^2}
\end{align}
and
\begin{equation}
     \F_{\theta_Z}^{(3)} = N \frac{\big(\left(1 + G_B \right)^{N-1} + \left(1 - G _B\right)^{N+1}\big) G''_B-\left(\left(1+ G _B\right)^{N-2} +\left(1 - G_B \right)^{N-2}\right) (G'_B)^2}{\left(1+ G_B \right)^N +\left(1 - G _B\right)^N}.
\end{equation}
Again we consider the two limits. In the thermodynamic limit $N\to \infty$ with fixed valued of $\beta, J, B$ the function $0< G_B <1$ is a positive constant between zero and one. In this limit we find
\begin{align}
    \F_{\theta_Z}^{(2)} &\to 0  \\
    \F_{\theta_Z}^{(2)} &\to N\left(\frac{G_B''}{1+G_B}  - \frac{(G_B')^2}{(1+G_B)^2}  \right) = N \frac{\partial^2} {\partial B^2} \log (1+G_B).
\end{align}
Combining with the last term one finds  $\F_{\theta_Z}=  \F_{\theta_Z}^{(1)}+\F_{\theta_Z}^{(2)}  = N \frac{\partial^2} {\partial B^2} \ln \left[ \cosh(\beta B)\right]$, which is straightforward to compute and gives 
\begin{align}
B>0\quad N\to \infty:\qquad  \F_{\theta_Z} = N \beta^2 \frac{e^{- 4 \beta J} \cosh (\beta  B)}{\left(\sinh ^2(\beta  B)+e^{- 4 \beta J}\right)^{3/2}}
\end{align}

Finally, in low temperature limit $\beta\to \infty$ at a fixed $N\gg 1$ and $B>0$, the function $G_B$ tends to $1$ from below, we thus write $G_B = 1-\delta_B$ and expand in small $\delta$. For the different contributions to the QFI we thus find
\begin{align}
    \F_{\theta_Z}^{(2)} &\to 0 \\
    \F_{\theta_Z}^{(3)} &\to N \left(\frac{G_B''}{2}- \frac{(G_B')^2}{4}\right).
\end{align}
Combining with $\F_z^{(1)}$ and taking the limit gives 
\begin{equation}
    B>0\quad \beta\to \infty: \qquad \F_{\theta_Z} = 4 N \beta^2 e^{- 2 \beta B}( e^{- 2 \beta B}+ e^{- 4 \beta J}).
\end{equation}


\end{document}